\documentclass[12pt,preprint,natbib]{aastex}

\RequirePackage{lineno} 
\usepackage{longtable}
\usepackage{epstopdf}  
\usepackage{ifthen}
\usepackage{rotating}
\usepackage{subfigure} 
\usepackage{booktabs}

\def\afrhot{{$A$($\theta$)$f\rho$}}
\def\afrho{{$Af\rho$}}
\def\deg{$^\circ$}
\def\rh{$r_{\mathrm{H}}$}

\begin{document}
\bibliographystyle{apj}

\setlength{\footskip}{0pt} 

\title{Observations of Comet ISON (C/2012 S1) from Lowell Observatory}

\author{Matthew M. Knight\altaffilmark{1,2,3}, {David G. Schleicher\altaffilmark{2}}}

\author{Submitted to {\it The Astronomical Journal}: July 31, 2014; Accepted: October 1, 2014}

\author{Manuscript: \pageref{lastpage} pages text (single spaced), \ref{lasttable} tables, \ref{lastfig} figures}

\altaffiltext{1}{Contacting author: knight@lowell.edu.}
\altaffiltext{2}{Lowell Observatory, 1400 W. Mars Hill Rd, Flagstaff, Arizona 86001, USA}
\altaffiltext{3}{Visiting scientist at The Johns Hopkins University Applied Physics Laboratory, 11100 Johns Hopkins Road, Laurel, Maryland 20723, USA}

\begin{singlespace}

\begin{abstract}
We observed dynamically new sungrazing comet ISON (C/2012 S1) extensively at Lowell Observatory throughout 2013 in order to characterize its behavior prior to perihelion. ISON had ``typical'' abundances for an Oort Cloud comet. Its dust production, as measured by {\afrho}, remained nearly constant during the apparition but its CN gas production increased by $\sim$50$\times$. The minimum active area necessary to support observed water production rates exceeded the likely surface area of the nucleus and suggests a population of icy grains in the coma. Together with the flattening of the dust radial profile over time, this is consistent with ejection of a large quantity of slow moving dust and icy grains in the coma at large heliocentric distance. The dust morphology was dominated by the tail, but a faint sunward dust fan was detected in March, April, May, and September. We imaged multiple gas species in September, October, and November. All gas species were more extended than the dust coma, although only CN had sufficient signal-to-noise for detailed morphological study. Excess CN signal was observed in the sunward hemisphere in September and early October. In November the excess CN signal was in the tailward hemisphere and two faint CN features appeared approximately orthogonal to the tail with position angles varying by about $\pm$20$^\circ$ from night to night. Using numerical modeling, we best reproduced the orientation and shape of these features as well as the bulk brightness with a pole oriented approximately towards the Sun and a single source located within $\sim$35$^\circ$ of the equator. Variations in position angle and relative brightness of the CN features from night to night suggest a rotation period shorter than 24 hr. The production rates and coma morphology suggest a nucleus that was active over nearly its entire sunward facing hemisphere in September and October but which underwent a significant mass loss event, potentially including fragmentation, shortly before November 1. Significant mass loss likely continued at the same site over subsequent days/weeks and may have catastrophically weakened the nucleus prior to perihelion. 
\end{abstract}


\section{INTRODUCTION}
Comet ISON (C/2012 S1) was discovered 2012 September 21 and was soon recognized as being both dynamically new and on a sungrazing orbit \citep{cbet3238}. This combination is unique; never before have we observed a comet newly arriving from the Oort Cloud reach such a small perihelion distance. Furthermore, all known sungrazing comets have been discovered inside of 1 AU and most are destroyed during the perihelion passage (cf. \citealt{sekanina02b,knight10d}), preventing detailed studies of their evolution with heliocentric distance. Thus, ISON represented an unprecedented opportunity to study a comet as it passed from beyond Jupiter, where activity is driven by hypervolatiles like CO and CO$_2$, through the zone of water-driven activity where comets are most frequently observed, and into the near-Sun region where dust and metals sublimate.

Recognizing this opportunity, we began observing ISON in 2013 January and observed it regularly through early November. These observations were intended to establish a baseline for comparison with its composition and behavior after perihelion. Since ISON did not survive perihelion (e.g., \citealt{knight14,sekanina14}) this paper represents the totality of our observations. In addition to its destruction near perihelion, ISON was fainter throughout most of its apparition than early expectations, so community data were somewhat limited. Published observational datasets have so far included characterization of the early dust activity \citep{meech13}, dust coma morphological studies at large heliocentric distance \citep{li13,hines14}, snapshot compositional analysis (e.g., \citealt{orourke13, shinnaka14, agundez14}), and observations from telescopes onboard Sun-observing spacecraft \citep{knight14,combi14,curdt14}.

The current work includes both compositional and morphological studies over a wider range of heliocentric distances than any papers on ISON yet published. Our observations and reductions are summarized in Section~\ref{sec:observations}. As we chose to concentrate our efforts on imaging, our photoelectric photometer observations (Section~\ref{sec:photometry}) were relatively limited and were supplemented with production rates derived from imaging. Nevertheless, they included the first published gas production rates \citep{iauc9254,iauc9257} as well as the first published detection of OH \citep{iauc9260}. The bulk of our imaging was snapshots owing to ISON's relative faintness throughout the first half of the year and the small nightly observing window once it reemerged from solar conjunction in August. Our observations of the evolving dust and gas coma morphology are discussed in Section~\ref{sec:imaging} while section~\ref{sec:modeling} summarizes the numerical modeling we conducted to explain the coma morphology. The implications of our observations are discussed in Section~\ref{sec:discussion} and conclusions are presented in Section~\ref{sec:conclusions}.

\section{OBSERVATIONS AND REDUCTIONS}
\label{sec:observations}

\subsection{CCD Observations and Reductions}
\label{sec:obs_ccd}
Imaging began 2013 January 12 on the inaugural science night for Lowell Observatory's 4.3-m Discovery Channel Telescope (DCT; \citealt{schleicher13}) and continued monthly to perihelion except when solar conjunction precluded it (July and August) or when we were weathered out (February). DCT images were acquired with the Large Monolithic Imager (LMI) which has a 6.1K$\times$6.1K e2v CCD with a field of view of 12.3 arcmin on a side. All LMI observations were binned 2$\times$2 on-chip, yielding a pixel scale of 0.240 arcsec. Four-inch diameter round HB narrowband comet filters \citep{farnham00} were used in September and October resulting in a partially vignetted field of view with a circular usable area approximately 12 arcmin in diameter. Full sized 4.75-in square Kron-Cousins broadband filters were used throughout and did not obscure the field of view. 

DCT observations were supplemented by imaging with the 42-in Hall telescope (1.1-m) from April onwards and by the 31-in telescope (0.8-m) in robotic mode in October and November. The 42-in images were obtained with an e2v CCD231-84 chip with 4K$\times$4K pixels and a field of view of 25.3 arcmin on a side. These were binned 2$\times$2 on chip yielding a pixel scale of 0.740 arcsec. The 31-in images were obtained with an e2v CCD42-40 chip with 2K$\times$2K pixels having a pixel scale of 0.456 arcsec and a field of view of 15.7 arcmin on a side. These observations are summarized in Table~\ref{t:imaging_circ}.

As shown in Table~\ref{t:imaging_circ}, most of our imaging consisted of snapshot observations. From January through April, when ISON was observable for much of the night, this was a deliberate choice due to its faintness, which restricted imaging to broadband filters, and due to ISON's observability window overlapping with our primary science target, 10P/Tempel 2, for which we sought to obtain a high temporal coverage lightcurve of its bare nucleus \citep{schleicher13}. While broadband B, V, R, and I images were acquired throughout the apparition, we emphasized R-band. We acquired narrowband CN and blue continuum (BC) images in April, but after removal of the underlying continuum (discussed below) there was no detectable CN signal. The comet's small solar elongation limited observations from May onward to less than 2 hr per night, primarily at high airmass and/or during twilight. Emphasis was placed on the narrowband comet filters when ISON became bright enough (September onward). While it was detected in all narrowband filters we used (see Table~\ref{t:imaging_circ}), only CN provided high enough signal-to-noise (S/N) for morphological assessment. All 31-in observations consisted of images acquired with broadband R or narrowband CN filters due to the small telescope size and limited observing window. All images were either tracked at the comet's rate or exposures were short enough that trailing was not an issue.

The bias was removed and a flat field was applied following standard reduction techniques. HB narrowband and broadband standard stars \citep{farnham00,landolt09} were observed on all photometric nights with the DCT or the 42-in. We used our standard photometric procedures \citep{farnham00} on these nights to determine flux calibrations and process the narrowband images into pure dust and pure gas images. We also flux calibrated CN images obtained with the 31-in on photometric nights by creating synthetic BC images from broadband R images, where the scaling between R and BC was determined from their ratios on photometric nights in which both were obtained. Since the signal from underlying dust in the CN filter was small (typically less than 20\%), any uncertainty introduced by the use of synthetic BC images was negligible for these purposes. On photometric nights gas features were distinctly different from dust features and were visible in both raw and decontaminated images (discussed in Section~\ref{sec:imaging}). Therefore, we were able to use contaminated CN images for morphological assessment on non-photometric nights.

To supplement our standard photometry discussed in the next section, we also extracted fluxes from subsets of CCD images acquired on photometric nights. Fluxes were converted to production rates and {\afrho} in the same manner as the photometer fluxes described next. The uncertainty in the CCD-derived fluxes was considerably higher than the photometer-derived fluxes primarily because no sky frames were acquired during imaging so sky values necessary for background removal were measured near the corners of the frames (and also due to bias and readout noise). ISON's gas coma extended to the edge of the CCD for all of our narrowband gas images but only contributed significantly in the DCT images, where the sky was measured closer to the nucleus due to the smaller field of view. The fluxes measured from 31-in images contained additional uncertainty due to our use of broadband R images and assumed extinction and instrumental coefficients to create synthetic underlying continuum images.

We determined the centroid of each image by fitting a two-dimensional Gaussian to the apparent photocenter. This was necessary for some image enhancement techniques and for processing into pure gas and dust images. As discussed in Section~\ref{sec:imaging}, the gas morphology did not affect the centroiding but dust images had a strong tail which may have biased the dust centroids slightly tailward. We applied various enhancement techniques (e.g., \citealt{schleicher04,samarasinha14}) to explore faint structures in the coma morphology. Numerous enhancement techniques were applied to confirm the validity of structures in the coma, but analyses discussed herein primarily utilized removal of an azimuthal median profile.

\subsection{Photometer Observations and Reductions}
\label{sec:obs_phot}
A traditional photoelectric photometer remains our standard instrument
for measuring gas and dust production rates and the associated observations
were obtained at the 42-in telescope in March, May, September,
and October (Table~\ref{t:phot_circ}). The HB
filter set was employed, but we only used two of our standard eight
filters (CN and blue continuum) in early 2013 due to the faintness
of the comet ($\sim$16$^{th}$ mag). These early data also required us to obtain
a long sequence of measurements ($>$1 hr) alternating between comet and sky
that were averaged to yield one resulting value per night. In other respects,
the photometry were acquired and reduced using our standard techniques
(cf.\ \citealt{ahearn95,schleicher11}) to fluxes, aperture abundances ($M$($\rho$)), and then
to production rates ($Q$) for each observed gas species -- OH, NH, CN, C$_3$,
and C$_2$ -- along with a vectorial-equivalent value for water based on OH.
The quantity {\afrho} \citep{ahearn84} was computed for the continuum filters, and as appropriate,
we also apply a phase adjustment of {\afrhot} to normalize to
0$^\circ$ phase angle (see \citealt{schleicher11}).
Because of unusually poor weather along with our priority of morphological
studies of ISON, standard photometry observations were only obtained
on four nights. Therefore, as noted in the prior subsection, we have
supplemented these data by extracting aperture fluxes from some of the
imaging data obtained on photometric nights. The nightly observing
circumstances for both data types are presented in Table~\ref{t:phot_circ}, along with
the associated reduction coefficients.

\section{PHOTOMETRIC RESULTS}
\label{sec:photometry}
\subsection{The Data Set}

We first present results from our various photometric measurements. 
In Table~\ref{t:phot_flux} we list reduced fluxes and aperture abundances, 
while the resulting production rates and {\afrhot} values are 
given in Table~\ref{t:phot_rates}. Formal photon statistical uncertainties 
are also provided in Table~\ref{t:phot_rates} 
for the photometer measurements, while estimated uncertainties are 
presented for the imaging data.

\subsection{Composition and Behavior with Time and Distance}
\begin{figure}
  \centering
  \includegraphics[width=175mm]{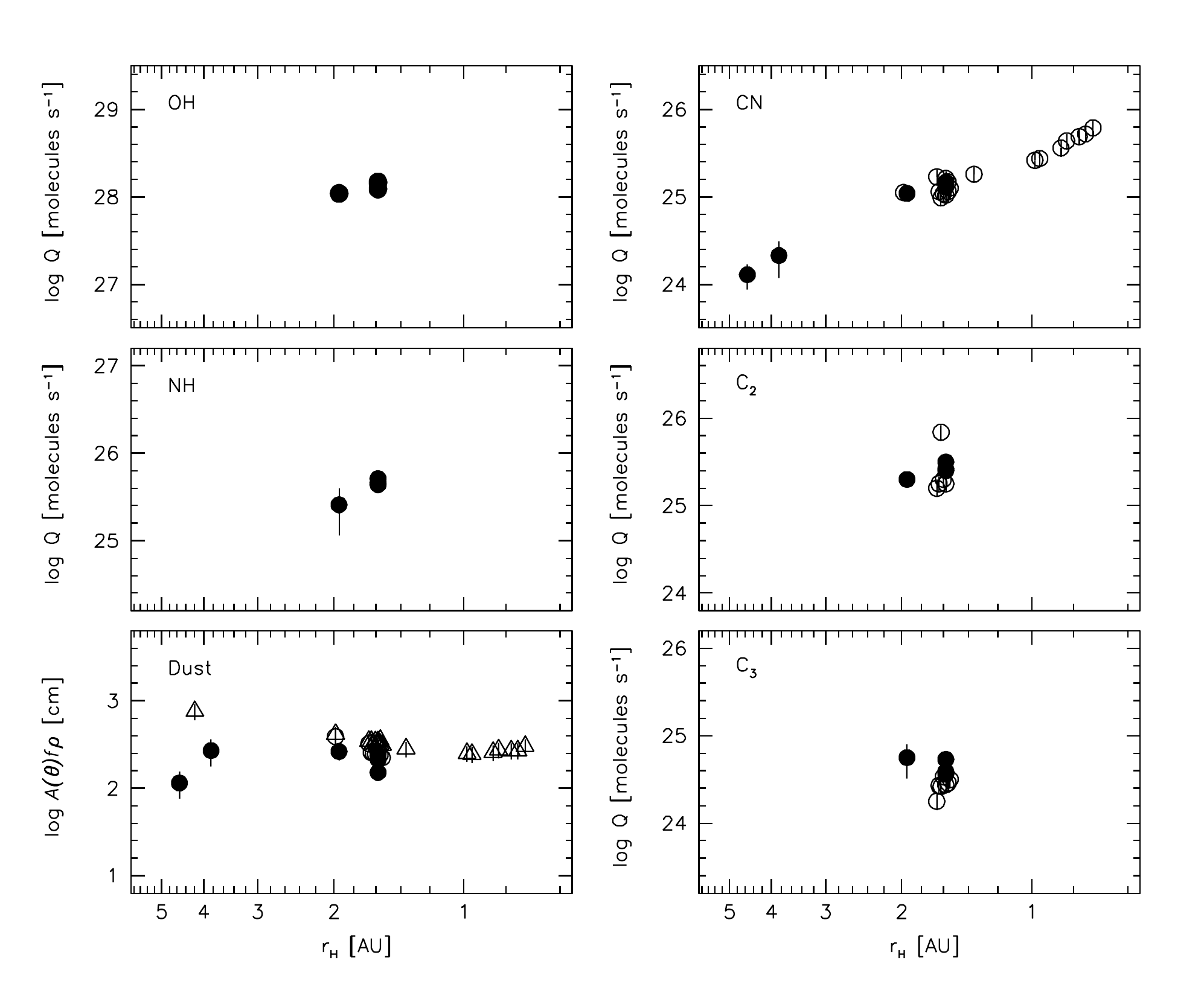}
  \caption[Log of production rates versus log of heliocentric distance]{Logarithm of the production rates for observed molecular species and $A(\theta)f\rho$ for the dust as a function of the heliocentric distance (plotted with logarithmic scale). The species is indicated in the upper left of each plot. In all panels filled points are photometer measurements while open points were determined from CCD imaging. In the {\afrhot} plot, circles are used for the blue continuum measurements while triangles were used for the R-band measurements. Error bars are plotted for all points; in some cases they are smaller than the symbols and are therefore not visible. Aperture sizes and other relevant parameters are given in Tables~\ref{t:phot_circ}--\ref{t:phot_rates}.}
  \label{fig:photometry}
\end{figure}

We plot in Figure~\ref{fig:photometry} the logarithms of the derived 
production rates as a function of the heliocentric distance on a logarithmic scale. Values extracted 
from images are distinguished from those derived from photometer 
measurements by open versus filled symbols, while R-band dust {\afrho} values 
from images are also identified by differing symbols from the blue 
continuum results.

Looking first at ISON's basic composition, abundance ratios among the 
gas species indicate that ISON had ``typical'' abundances, as would be 
expected for an Oort Cloud origin (cf. \citealt{ahearn95,bair12}), 
but the ratios of the minor species to OH 
are all near the low end of their respective typical ranges.
Note that this is not associated with its age, as dynamically new 
comets encompass the full typical range. 
Unexpectedly, our only night with multiple apertures with the 
photometer (October 4) revealed significant and systematic trends with 
aperture size for the gas species. These ranged from 15\% for CN to 47\% 
for the short-lived C$_3$, with the largest aperture yielding the 
highest production rates. Although fewer species were measured using 
photometry extracted from images, these cases also exhibit similar trends.
The only previous times that we have observed such persistent trends 
with aperture size were for very high production rate comets 
such as 1P/Halley and Hale-Bopp (C/1995 O1) when the radial outflow 
velocites were systematically higher than normal \citep{schleicher89,schleicher99}, 
and for Comet 17P/Holmes, where the outgassing rate dropped 
systematically following its epic outburst. In the latter case, 
an expanding icy grain halo coupled with a strong decay in the rate of 
outgassing produced non-equilibrium conditions in Holmes \citep{schleicher09}.
For ISON, we suspect that the observed aperture trends might also be 
caused by an expanding halo of residual icy grains released early in the 
apparition. Given the long duration, we suspect these grains must have been 
relatively large in size with very low velocities and/or the halo was
replenished by the ongoing release of additional icy grains.

\begin{figure}
  \centering
  \includegraphics[width=88mm]{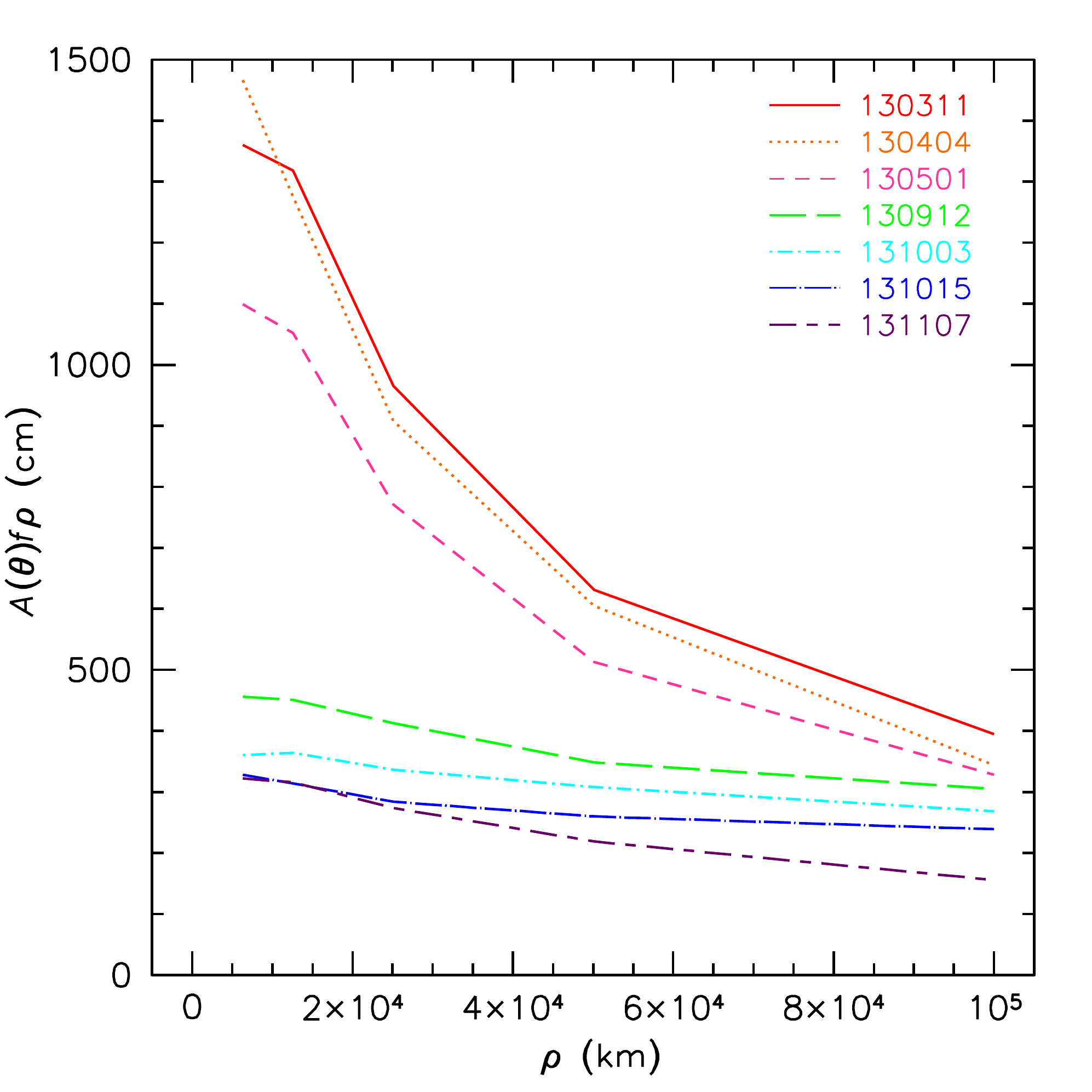}
  \caption[images]{R-band {\afrhot} as a function of distance from the nucleus ($\rho$) during the apparition. One representative curve is plotted for each observing epoch. The specific night for each run is identified in the key. The overall trend is for a decreasing and flattening of {\afrhot} over time. Note that it was not photometric on March 11, April 4, or May 1 so the absolute calibrations may not be accurate but the trend is unaltered. On March 11 and April 4 the sky was reached slightly before the largest aperture plotted so the slope between the two largest apertures is slightly steeper than in reality these months. On all other nights coma was detected beyond 100,000 km.}
  \label{fig:afrho}
\end{figure}

Despite this behavior for the gas species in the fall, {\afrho}
exhibited the opposite aperture trend throughout the apparition
caused by a steeper radial fall-off in the dust than the canonical 1/$\rho$
where $\rho$ is distance from the nucleus. 
The departure from 1/$\rho$ was greatest early in the 
apparition and was much less by the fall, as demonstrated 
in Figure~\ref{fig:afrho} where we show {\afrho} for varying aperture sizes at several 
representative times. 
The decreasing departure from 1/$\rho$ during the apparition is not the result of changing coma size. While {\it Hubble Space Telescope} ({\it HST}) observations in April reported a ``well-defined'' coma in the sunward direction extending out to only $\sim$9000 km \citep{cbet3496}, we detected coma out to 70,000--80,000 km in March and April, and beyond 100,000 km in May. Thus, at most our largest aperture plotted in Figure~\ref{fig:afrho} was biased downward by $\sim$25,000 km of empty sky in March and April. This is a $\sim$25\% effect that is considerably less than the factor of $\sim$3 decrease in {\afrho} from the smallest aperture during those months (as compared to a decrease from the smallest to largest aperture of only $\sim$1.5 in the fall).

The severe steepness in the spring explains the 
apparent large increase in {\afrho} from March to May as being caused 
by the use of a smaller aperture in May (to avoid a nearby star). 
It also explains the apparent discrepancy between our large aperture 
photometer results \citep{iauc9254,iauc9257}
and the very small aperture result reported from 
{\it HST} observations in April \citep{cbet3496} along with our April 
imaging value (Figure~\ref{fig:photometry}) which is at an intermediate 
aperture size.
Since the comet did not brighten appreciably during the spring, 
we cannot attribute this strong aperture trend to increasing 
activity. Instead, we suspect that very low velocity grains released 
near or prior to the beginning of 2013 remained in the inner coma 
throughout the spring; such grains must have had outflow velocities 
in the meters per second range, consistent with recent modeling of
dust in the coma of comet Siding Spring (C/2013 A1) at comparable
heliocentric distances \citep{tricarico14}.

To examine the evolution of dust production with heliocentric distance, 
we have extracted values for a constant ${\rho}=25,000\ \mathrm{km}$ 
from the images -- an aperture size for which seeing effects are 
negligible and S/N is maximized -- and used the curves from images 
such as shown in Figure~\ref{fig:afrho} to extrapolate the photometer measurements 
to the same projected aperture radius for Figure~\ref{fig:water_rates}.
To increase our temporal coverage of {\afrho}, we have also made use
of our R-band imaging. Because our and other data sets indicate 
the dust in ISON exhibited little or no reddening compared to solar 
color \citep{li13,cbet3598,cbet3608},
we have simply included our R-band results with the blue continuum 
results in Figures~\ref{fig:photometry} and \ref{fig:water_rates}, making no adjustments for wavelength as 
the color terms are smaller than the size of the plotted symbols. 
Finally, Figure~\ref{fig:water_rates} includes an adjustment for phase angle 
(cf.\ \citealt{schleicher11} and references therein) -- normalizing
to 0$^\circ$ -- since the 
comet's phase angle varied from 11$^\circ$ to 70$^\circ$ during these observations 
(see Table~\ref{t:phot_flux}).

As evident from Figure~\ref{fig:water_rates}, dust production as defined by $A(0{^\circ})f{\rho}(25,000\ \mathrm{km})$ 
was nearly constant throughout our observing interval, even as gas production 
as evidenced from CN in Figure~\ref{fig:photometry} increased by a factor of 50. Note that 
the overall {\rh}-dependence of $Q$(CN) in log-log space was $-$1.9, somewhat 
steeper than previously found for a pair of inbound, dynamical-new 
comets by \citet{ahearn95}, while this same pair of objects 
also had near-constant dust production. While the dust-to-gas ratio 
varies by over two orders of magnitude among comets, the dust-to-gas 
ratio for a given comet as a function of heliocentric distance seldom 
varies by more than a factor of a few; as a class, inbound, dynamically 
new comets appear to be quite different. Since these comets are known 
to have unusually high activity for their respective nucleus size at 
large {\rh}, we suggest that a population of large, slow moving grains 
are released at large {\rh} and remain in the coma as the comet approaches 
the Sun, dominating the {\afrho} measurements. We will return to this 
possibility in Section~\ref{sec:discussion}.

While our own observations ended earlier than planned due to an unfortunate 
sequence of a dome failure and then an early winter storm system, other investigators 
report a large outburst in ISON occuring between November 11 and 13, with 
{\it both} gas and dust increasing by about a factor of 12 \citep{cbet3711c}.
Looking at our sequence of dust and gas  measurements in early November, 
it appears that the $\sim$15\% increase on November 11.5 as compared 
to the previous 
nights indicates the onset of the outburst. Further, we argue that the 
similar increase in dust and gas production directly implies that only 
with the outburst does newly released dust finally dominate over old 
grains released much earlier in the apparition.

\subsection{Water Production and Active Area}

As the comet was only bright enough to cleanly measure OH in the 
fall with the photometer (even with the DCT, the net count per pixel 
for imaging was dominated by readout and bias uncertainties), 
our direct OH measurements are insufficient for examining the 
variation throughout the apparition. However, we can supplement 
these observations 
by assuming that the OH-to-CN abundance ratio was constant throughout 
the apparition. We consider this to be a reasonable assumption in the 
early fall, since we measured the identical value -- a log abundance ratio 
of 3.00 -- in September and early October. However, some comets have 
exhibited a progressive increase in the CN-to-OH ratio inside 1 AU 
believed to be an artifact from the Haser model \citep{ahearn95}, and 
we estimate that OH might be correspondingly lower 
than we assume by as much as 40\% on our last night (this trend is 
consistent with preliminary analysis of optical spectroscopy of ISON 
by \citealt{mckay14}). 
The validity of the measured ratio to computing values in the spring 
based on CN is even less certain, as we do not know if outgassing 
at this time was dominated by water vaporization, or if icy grains 
containing water and/or the CN parent were released by more 
volatile species such as CO.

With these cautions in mind, we have used our measured CN production 
rates and the measured ratio of 1000 to compute OH $Q$s throughout the 
apparition. As described in Section~\ref{sec:obs_phot}, these were then used to 
compute vectorial equivalent water production rates, which are 
plotted in Figure~\ref{fig:water_rates}. As there is a {\rh}$^{-0.5}$ factor 
in the vectorial model not present in our Haser model, the overall 
slope from our data is $-$2.4, rather than the CN value of $-$1.9.

We also plot in Figure~\ref{fig:water_rates} other water production rates 
made with a variety of 
techniques and reported by different teams of investigators (see key). 
As is evident, we are in generally good agreement with the ensemble 
of measurements. Error bars are not shown, both for clarity and because 
some reported measurements did not include uncertainties. 
Taken together, there is clear evidence for a more shallow slope 
between 2 and 1 AU than the apparition as a whole, especially as the 
values at large {\rh} based on CN are likely higher than the true water 
production. The previously mentioned outburst that apparently began on November 11 
($r_\mathrm{H}=0.72$ AU) is also clearly evident, and appears to have 
continued through the last reported water measurements with SWAN on 
November 23.6 at $r_\mathrm{H}=0.32$ AU \citep{combi14}, possible evidence 
that ISON was beginning to break apart.

Making some standard assumptions regarding vaporization of ice from 
the surface of a cometary nucleus (cf.\ \citealt{cowan79}), we have computed the minimum active 
area required to yield the observed water production, and these 
results are presented in the bottom panel of Figure~\ref{fig:water_rates}. The near-level 
values between 1.3 and 0.7 AU are likely representative of the true 
active area on the surface -- $\sim$1.5 km$^2$ for the sub-solar case or 
$\sim$6 km$^2$ for the uniformly active, isothermal case -- while we think 
that the higher values prior to this interval also contain a component 
from vaporizing icy grains in the coma. 
The large increase with the outburst on November 11 is either due to fragmentation of 
the nucleus or ongoing release of large quantities of fresh icy grains.

As usual for a new comet, the nucleus was never detected due to being 
overwhelmed by light from the coma. The only published upper limits were 
reported by \citet{cbet3496} from HST imaging in April (effective nucleus radius $R_{N} <2$ km) 
and by \citet{cbet3720a} from the Mars Reconnaissance Orbiter's 
HiRISE instrument in late September -- early October ($R_{N}<0.6$ km).  For a uniformly 
active surface, as one would expect for a comet approaching the Sun 
for the very first time, this later upper-limit implies that $\sim$100\% 
of the surface was, indeed, active. Even the sub-solar solution for 
effective active area implies at least one-quarter of the surface was 
active. Thus it seems very unlikely that ISON  only had an isolated source 
region on its surface as suggested by some investigators.

\section{SPATIAL DISTRIBUTIONS AND MORPHOLOGY}
\label{sec:imaging}
\subsection{Dust Distribution and Morphology}
\label{sec:dust_morph}
\begin{figure}
  \centering
  \includegraphics[width=75mm]{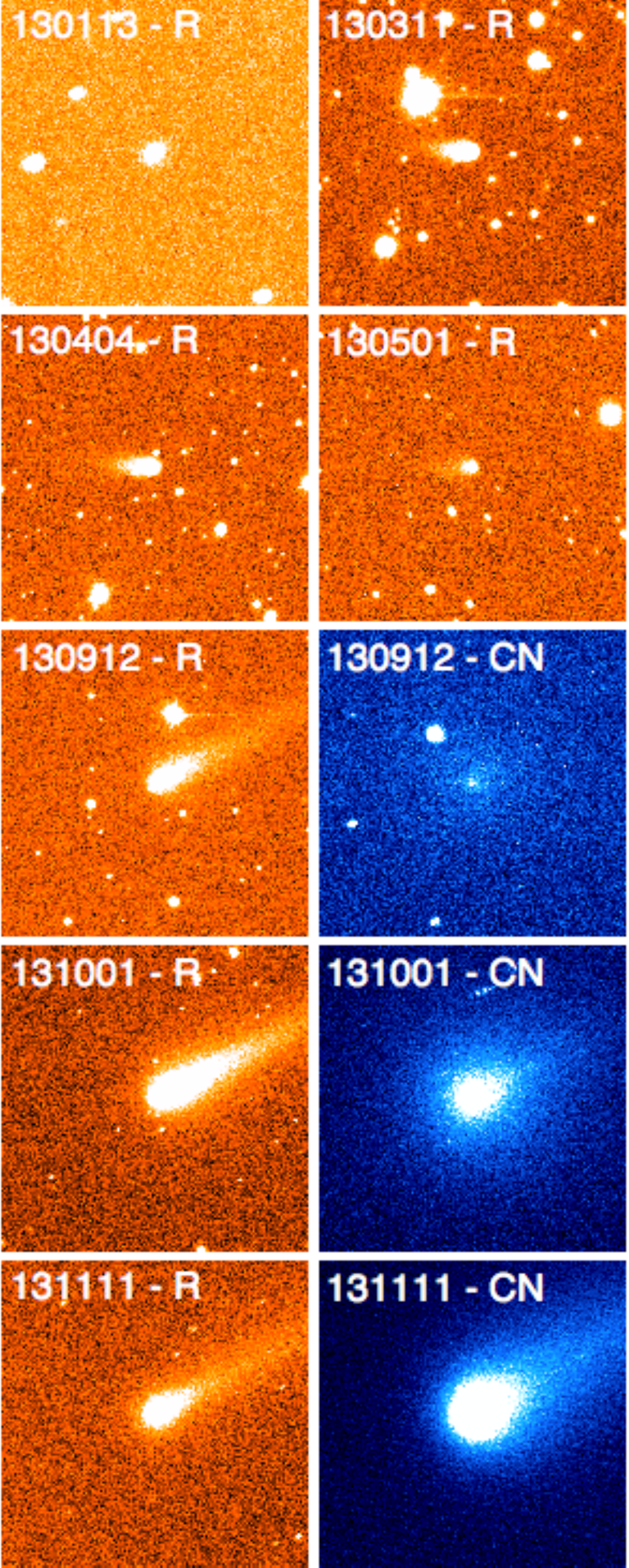}
  \caption[images]{Evolution of bulk morphology. The date (YYMMDD format) and filter of each image is shown. Each image is 500,000 km on a side and is centered on the comet. R-band images have an orange/white color scheme and CN images have a blue/white color scheme. The color stretches vary from panel to panel, but white is brightest and black is faintest.}
  \label{fig:morph}
\end{figure}

We now turn our attention to imaging, which will be the focus of the remainder of the paper. Figure~\ref{fig:morph} shows representative images of the evolution of ISON's appearance in broadband R images. R-band images are relatively free of gas contamination, as we confirmed when gas images were obtained, and are therefore dominated by dust reflecting the solar continuum. When first observed in January, the comet appeared diffuse with minimal tail evident. The visible extent of the dust tail increased as ISON brightened, eventually extending beyond our field of view in October ($\sim$300,000 km in the plane of the sky).

\begin{figure}
  \centering
  \includegraphics[width=88mm]{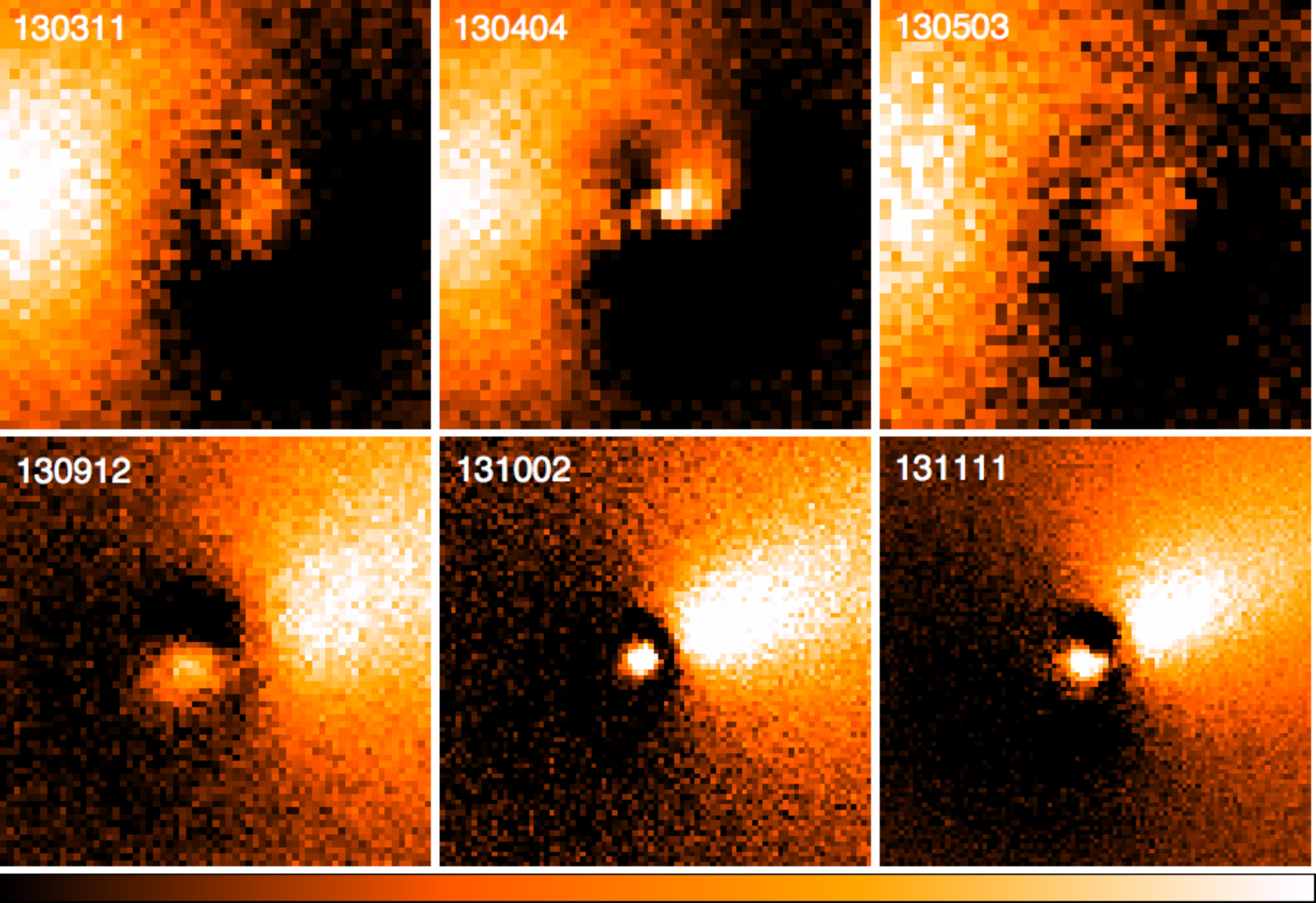}
  \caption[images]{Evolution of the sunward dust feature. Each image is centered on the nucleus, is 30,000 km across at the comet, and has been enhanced by subtraction of an azimuthal median profile. The date of each image is given on each panel (YYMMDD format). The color stretches are different in each panel, but in all cases white is brightest and black is faintest. The Sun is at a PA near 270$^\circ$ in the top row and near 110$^\circ$ in the bottom row (specific PAs for each night are given in Table~\ref{t:imaging_circ}).}
  \label{fig:dust_feature}
\end{figure}

Multiple image enhancement techniques revealed a feature roughly in the sunward direction in R-band images acquired throughout much of the apparition (Figure~\ref{fig:dust_feature}). During March, April, and May the feature extended 6000--7000 km before turning around towards the north and moving tailward. No feature was detected in January or June, but the the lack of appearance during these months is not conclusive owing to the comet's faintness (January) and extremely high airmass (June). The feature was nearly straight in September, extending radially $\sim$5000 km. A sunward feature was seen in October and November but was shorter in extent than other months (2500--3000 km) and its appearance and position were highly sensitive to the centroid position. Thus, we suspect that it was not real in October or November, but was an artifact of the enhancement process since the bright tail may have biased the centroid slightly in the anti-sunward direction, yielding a small amount of excess signal in the sunward direction.

Despite the relatively small extent of the dust feature and its low contrast relative to the coma, we consider it likely to be real in March, April, May, and September for several reasons. It was 1.5--3$\times$ greater in extent than the effective seeing in every month except March, when it was roughly comparable to the seeing, which was extremely poor. The feature was seen in each individual frame during these months, and was similar in appearance when the centroid was shifted by 1 pixel in each direction. The dust feature matches the PA (position angle), shape, and extent of the dust feature discovered in {\it HST} images acquired on April 10 \citep{cbet3496,li13} and seen in {\it HST} images acquired May 7 \citep{hines14}. The relatively small change in geometric circumstances and the similar morphology between March, April, and May argue strongly in favor of its existence in March as well. The feature had its greatest extent both in pixels and relative to the seeing during September, making it unlikely to be an artifact of the processing. Our conclusion that the October and November features are artifacts of the processing are consistent with non-detections of a dust feature in {\it HST} images acquired on October 9 and November 1 (J.-Y. Li, private communication, 2014).

The dust feature did not appear to change position or shape during a night or from night to night. It was centered at a PA of 270$^\circ\pm$20$^\circ$ in March, 300$^\circ\pm$10$^\circ$ in April, 250$^\circ\pm$20$^\circ$ in May, and 120$^\circ\pm$15$^\circ$ in September, having an opening angle of $\sim$60$^\circ$ each month. Since no changes in the morphology were seen over $\sim$19 hr of {\it HST} observations, \citet{li13} proposed that the dust originated from a polar source. Under this assumption, the PA of the middle of the feature indicates the comet's projected pole, and each epoch provides a great circle of possible pole solutions. Using our dust PAs from March, April, May, and September, we find a pole at R.A. = 296$^\circ$$\pm$5$^\circ$ and Dec = $-$27$^\circ$$\pm$5$^\circ$. This is $\sim$10$^\circ$ from one end of \citet{li13}'s preferred range of solutions, but is more robust since our PAs were measured across a range of viewing geometries, thus constraining the solution.

However, we believe that this solution for the rotation axis is unlikely for several reasons. During the fall this pole pointed progressively closer to the plane of the sky, so the dust feature should have become more obvious and well defined; instead, the dust feature was shorter and less well defined in September than in April, and was not seen at all in October or November. Second, if this was the only source of activity then all of the observed water and CN should also be traceable to it. As previously noted, the water production rate implies that at least one quarter of the surface was active -- hardly a small source -- while a source at this location does not replicate the CN features seen in November (discussed in the next subsection). Third, from a statistical standpoint, we find it improbable that ISON was so fortuitously aligned, having its pole directed towards the Sun and its only active region also located at that pole. 

We propose that the dust feature was due to enhanced activity at the subsolar point, as has been suggested for other comets at similar distances, e.g., Hale-Bopp (C/1995 O1) at 6 AU \citep{weaver97}, Christensen (C/2006 W3) at 5 AU \citep{valborro14}. In this scenario, the shortening of the feature in September and its disappearance in October and November can be explained by decreasing contrast of the feature relative to the overall coma as the comet approached the Sun and sublimation increased across the surface. Alternatively, the feature could have been produced by an extended outburst (in either the polar or sub-solar source scenario) that increased in activity between March and April and decreased thereafter. This would explain the relative clarity of the feature in April relative to other months despite the same telescope, comparable seeing, and comparable relative brightness (except in September when ISON was brighter). Additional, less likely, possibilities include that the feature seen in September may have been entirely unrelated to that seen in the spring, or that the nucleus was precessing so the PAs of the dust feature did not actually constrain the pole solution.

\subsection{Gas Morphology}
We successfully imaged ISON's gas coma from September onwards. As expected, the spatial extent of the gas coma was considerably larger than the dust coma (Figure~\ref{fig:morph}). While narrowband imaging was obtained for OH, CN, C$_3$, C$_2$, and CO$^+$ on various nights (Table~\ref{t:imaging_circ}), only CN yielded sufficient signal to permit detailed investigations of its morphology via image enhancement techniques. Thus, the remainder of this subsection deals only with CN.

\begin{figure}
  \centering
  %
  \includegraphics[width=88mm]{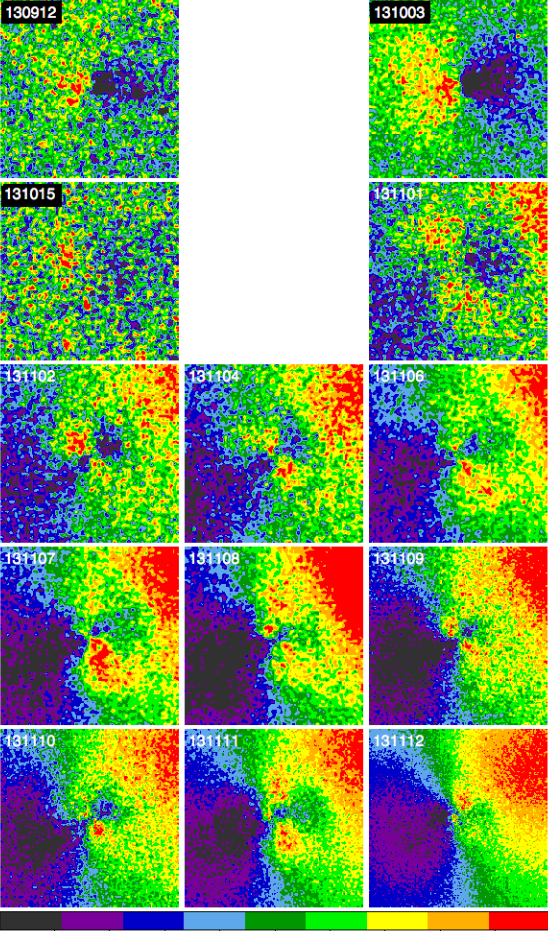}
  \caption[images]{Evolution of CN coma morphology. The date is given in the top left corner of each panel (YYMMDD). Each image is centered on the  nucleus, is 60,000 km across at the comet, and has been enhanced by subtraction of an azimuthal median profile. Gaussian smoothing with a radius of 3 pixels was used Sep 12--Nov 8, with a radius of 2 pixels on Nov 9--11, and with a radius of 1.5 pixels on Nov 12. The color stretch is different from image to image but the same color table is used for all images. Images on non-photometric nights (Nov 4, 8, 12) are contaminated CN; images on all other nights are decontaminated CN. Blank panels separate different months. The Sun is at a PA near 110$^\circ$ in all images (specific PAs for each night are given in Table~\ref{t:imaging_circ}).}
  \label{fig:gas_enhanced}
  \label{lastfig}
\end{figure}

The CN coma was nearly spherically symmetric in September and October but a faint, broad brightness excess in the sunward hemisphere was discernible in the inner coma after image enhancement (Figure~\ref{fig:gas_enhanced}). This suggests that the nucleus was active over much or all of the sunward hemisphere, as expected for a dynamically new comet and consistent with our interpretation of the water production rates. At larger distances from the nucleus a brightness excess in the tailward region of enhanced images first appeared in late-September and continued through November, when it was significantly brighter. This was likely due at least in part to an extended source of CN that was subject to radiation pressure, such as icy grains.

Two CN features roughly 180$^\circ$ apart and orthogonal to the dust tail appeared in our November images. While the S/N of these features was relatively low, both features were seen on every night we obtained data from November 1--12 and were evident before and after continuum was removed. The relative brightness of the features varied greatly from night to night as can be seen in Figure~\ref{fig:gas_enhanced}, where the relative intensities of the northern and southern features were similar on November 1, the southern feature was much stronger on November 7, and the northern feature was much stronger on November 12. The PAs also varied from night to night and we list them in Table~\ref{t:cn_pas}. However, no obvious variations in the shapes or brightness were seen during the limited observing window each night. No corresponding features were observed in dust images on any of these nights.

These features require enhanced activity from one or two regions on the nucleus, from which the CN parent is released. Under the assumption that the varying morphology was periodic and tied to the nucleus' rotation, we tried to constrain the rotation period by phasing the relative strength and PA of the CN feature on each  night. This yielded potential periods of 8.9, 10.4, 11.4, 12.6, 14.2, or 18.4 hr. Shorter periods were ruled out due to the lack of obvious change in the PA of the CN feature during a given night's observations. Longer periods were ruled out because there was evidence for both the northern and southern features in every image. The observed extent of the features in the plane of the sky was 15,000--20,000 km from the nucleus so, for a gas outflow velocity $\sim$1 km s$^{-1}$, gas took $\sim$6 hr to traverse the feature, and significantly longer if highly projected. Since the CN feature tracks the rotation of the source, if the source had been pointing away from a given hemisphere for more than 6 hr and was in the plane of the sky, the CN should have completely traversed the feature without additional CN following behind it, but material moving mostly toward or away from us could remain visible for considerably longer periods. This potentially explains the strong brightness variations between hemispheres from night to night and the presence of both features on all nights, but cannot constrain the rotation period further without knowing the degree of projection.

Weaver et al.\footnote{http://isoncampaign.org/observation-logs\#20131101} found a period of $\sim$10.4 hr for a single peaked lightcurve based on seven inner coma brightness measurements by {\it HST} on November 1 and noted that ``periods within the range 8--12 hr give acceptable fits to the observations.'' The signal measured by {\it HST} was dominated by dust in the inner coma rather than the nucleus, thus variations in brightness were caused by changes in activity, not changes in the apparent cross section of the nucleus. Assuming that the gas flow drops when a source region is not receiving sunlight, less dust would be entrained during local ``night.'' For one source region this would create a single peaked sinusoidal lightcurve, while for two source regions this would create a double peaked sinusoidal lightcurve. If the lightcurve was double peaked, Weaver et al.'s preferred period would be $\sim$20.8 hr, which is inconsistent with the variations in CN morphology we observed. As will be discussed in the following section, our modeling favors a single source solution, thus supporting the Weaver et al.\ results and implying one of our shorter periods is the most likely rotation period.

The first appearance of CN features by November 1 was likely due to an intrinsic  change in the comet and not simply due to the improvement in the S/N as it brightened. Accounting for the increase in the CN flux between our final night of DCT observations on October 4 and our November 1 31-in observations and the differences in telescopes (collecting area, pixel scale, quantum efficiency), the effective S/N was comparable on our final DCT run and the beginning of the November 31-in run. Thus, if the CN features were present with a comparable contrast to the ambient CN coma by October 4, they should have been detectable in our earlier images. It is unlikely that they appeared prior to October 15, as the effective S/N was only $\sim$2$\times$ lower on that night than November 1, so some hint of the features would likely have been visible. Thus, we conclude that these features likely originated in the second half of October. This correlates with the bulk brightness flipping from the sunward to the tailward hemisphere and suggests that the tailward bulk brightness observed in November was also tied to the emergence of the CN features.

This timeline is consistent with published descriptions of the coma morphology by other observers. 
The TRAPPIST team first clearly detected features in CN images on October 31, their first data since October 19 (Opitom, private communication, 2014). They later detected the features in C$_2$, OH, and dust continuum filters at the same PAs as the CN features \citep{cbet3693a,cbet3711c}\footnote{Opitom (private communication, 2014) notes that there was an error on CBET 3693, and their measured PAs were actually 10$^\circ$ and 190$^\circ$, not 10$^\circ$ and 90$^\circ$ as stated.}. Similar morphology was reported in broadband Bessell R \citep{cbet3715} and unfiltered \citep{cbet3718b} images acquired on November 14, and was reportedly not present on similar images acquired by both groups on November 13.

\section{MODELING}
\label{sec:modeling}
In an effort to better understand the cause of ISON's morphological features, we conducted numerical modeling with our Monte Carlo jet modeling code (e.g., \citealt{schleicher03a}). The code utilizes 10$^5$ particles to simulate cometary activity, with changes to any of numerous parameters being reflected in an updated model nearly instantaneously. This permits us to quickly explore a large variety of parameters including orientation of the nucleus' pole, source location(s) and extent(s), outflow velocities, parent/daughter lifetimes, position in the orbit, etc. Due to the large number of free parameters in the model and ISON's minimal observational constraints, we focused on replicating the gross coma morphology by constraining the pole orientation and varying the location(s) of active regions on the surface. For all modeling we have assumed ISON is a simple rotator; our Monte Carlo code can simulate non-principal axis rotation but, in the absence of data suggesting this, doing so would needlessly complicate matters.

As discussed in Section~\ref{sec:dust_morph}, we determined a pole solution from the assumption that the sunward dust feature originates from a near-polar source region. However, our modeling reveals that a source at this location cannot replicate the CN morphology observed in November as this would be approximately orthogonal to what we observed. Instead, we proposed that the dust feature was produced by activity from the subsolar point. Since this means that the dust feature cannot be used to constrain ISON's pole orientation, we investigate below two methods of replicating the CN features seen in November: 1.)\ two source regions, one located near each pole, and 2.)\ one source region near the equator.

In the two source region scenario, each source region must have been located near the rotation pole since the PAs of the CN feature changed little from night to night. Sources farther from the pole would have swept out a corkscrew pattern as seen in other comets, e.g.\ C/2007 N3 Lulin \citep{knight09b}. Thus, the pole is constrained to a great circle by the midpoint of one feature (the midpoint of the feature in the opposite hemisphere yields the opposite pole). The ensemble of observations from November 1--12 yields similar great circles since the viewing geometry changed relatively little during the interval, so we used the average to define the pole. We found that sources within $\sim$15$^\circ$ of each pole produced features with similar shapes to what was observed. We then stepped through possible pole solutions in 10$^\circ$ intervals along the great circle, modeling the resulting coma morphology between November 1--12 for each potential solution.

This scenario can match the changing PAs of the CN feature from night to  night and can explain the variations in brightness as due to the source regions being located at different longitudes. However, it cannot explain the bulk brightness enhancement in the tailward hemisphere being aligned with the midpoint of the CN features. Material leaving the nucleus from the poles would be expected to produce brightness enhancements in the north and south, not along the midpoint to the west. A second, more glaring, issue with this scenario is the simultaneous appearance of these features in late October as well as the increase in contrast of both features following the outburst on November 11--13. It is highly improbable that two source regions located near opposite poles would turn on at the same time and even less probable that they would later experience simultaneous outbursts of similar strength.

Our second and preferred scenario involves a single near-equatorial source
sweeping out a spiral that was seen nearly edge on. In this scenario, the resulting feature appeared clearly when it was in the plane of the sky (orthogonal to our line of sight) but was difficult to distinguish from the ambient coma when pointing towards or away from us (along our line of sight). The rotation axis would be near the plane of the sky and the direction of the pole is indicated by the angle that bisects the north and south feature PAs. 
We explored possible source locations for 10$^\circ$ steps 
along this great circle finding that solutions within $\sim$35$^\circ$ of the equator produced two features in approximately the correct orientation. However, the PA of the features and the location of the bulk brightness to the west
were best reproduced for a relatively small range of solutions near R.A. = 89$^\circ$$\pm$10$^\circ$ and Dec = $+$30$^\circ$$\pm$7$^\circ$.

Even more than the two source solution, this readily explains the PA variation and the bulk brightness variation. However, unlike the two source solution, this solution also naturally explains the bulk tailward brightness enhancement being aligned with the midpoint of the features. Material released when the jet is pointing towards or away from the Earth will be near the middle, increasing the amount of material seen in the west as compared to the north and south. This scenario also easily explains the surge in brightness of both features following the November 11--13 outburst since increasing activity from the single source region would affect both the north and south features. As discussed previously, a single source is also preferred when comparing the variability in ISON's inner coma brightness seen by $HST$ with our possible rotation periods based on the relative brightness and PAs of the CN features. It is possible that inclusion of as yet unpublished data from other observers could allow the rotation period to be conclusively determined, with the progression of the PAs of the feature yielding the sense of rotation (prograde or retrograde).

This pole solution is $\sim$30$^\circ$ from the pole inferred from the dust feature, which we have already discounted. With the dust pole, any choice of source locations will produce a spiral quite different from the CN features we observed in November. The $HST$ pole solution magnifies the problem, producing even more extreme spirals for any source locations, and therefore also not matching the CN morphology.

\section{DISCUSSION}
\label{sec:discussion}
As introduced earlier, the minimum active area necessary to support the measured water production rates between 1.3 and 0.7 AU was $\sim$1.5 km$^2$. This is approximately the cross section of sunlight that an $R_N=0.6$ km nucleus intercepts, e.g., it is consistent with ISON's sunward hemisphere being nearly 100\% active for published upper limits of the nucleus size \citep{cbet3720a}. Earlier in the apparition, the minimum active area was considerably higher, greatly exceeding the nucleus' surface area. Thus, we concluded that there was likely a substantial population of icy grains providing the needed surface area. Outgassing from an icy grain halo was predicted by \citet{huebner66} and has been suggested for other comets (e.g., Comet Bowell 1980b by \citealt{ahearn84}; 73P/Schwassmann-Wachmann 3 by \citealt{fougere12}; C/2009 P1 Garradd by \citealt{paganini12,villanueva12,combi13,bodewits14}) and conclusively demonstrated for 103P/Hartley 2 \citep{ahearn11a,kelley13,knight13,protopapa14}. Icy grains have already been suggested by other authors to explain ISON's early brightness behavior \citep{meech13}, the blue color of ISON's dust near the nucleus in April \citep{li13} and the lower negative-polarization of ISON's circumnuclear halo in May \citep{hines14}.

In this scenario, a large population of icy grains was present in ISON's coma early in 2013 but gradually diminished until it contributed negligibly to the total outgassing rate by late October. This would explain the disappearance of the bluer dust near the nucleus in the October 9 {\it HST} observations as compared to the April 10 observations \citep{li14} and is consistent with preliminary analysis of the October 26 {\it HST} polarization observations (D. Hines, private communication, 2014). One possible explanation for the origin of such a population is that they were released in a single event at large heliocentric distance, such as during the CO outburst suggested by \citet{meech13}. If released in such an event, the icy grains would have needed to be large enough to survive until near $r_\mathrm{H} = 1$~AU, implying that they had radii of 10s of cm if they were dirty ice \citep{beer06}. We see evidence for such a population of slow moving (velocities of order 1 m s$^{-1}$) grains in the flattening of our {\afrho} profiles during the apparition. Alternatively, a relatively low flux of smaller dirty ice grains (e.g. 10s of $\mu$m to $\sim$1 cm size) could have been released nearly continuously from large heliocentric distance inward. At large distances these grains would have survived for days to weeks, but at smaller distances would have only lasted minutes to hours \citep{beer06}.

The presence of a population of icy grains explains why the CN and water production rates did not increase at a constant rate (in log-log space) but flattened significantly as ISON approached the Sun. The nearly flat production rate from September to early November likely occured as lingering icy grains, which had dominated activity early in the apparition, finally left our photometic apertures and/or the increasing insolation shortened their lifetimes and they were destroyed faster than they were replenished by new production from the nucleus. It was only with the significant outburst November 11--13, when gas production rates increased by 12$\times$ \citep{cbet3711c} that sufficient quantities of new material were being produced to dominate over lingering material.

The sudden appearance of CN features in November, several months after ISON crossed the ``snow'' line where water ice sublimation begins to dominate activity (e.g., \citealt{meech04}) followed by a rapid increase in overall gas production is reminiscent of the seasonal behavior of Jupiter family comets. 
This behavior has been interpreted as being due to a loss of volatiles across much of the surface, resulting in activity being confined to a few isolated source regions (e.g., \citealt{ahearn95}). Until these regions are exposed to direct sunlight, they are inactive, with activity typically increasing steeply once illuminated. 
It is unlikely that ISON's surface was highly evolved; its relatively large brightness at large heliocentric distance and subsequent slower than expected increase in brightness is common for dynamically new comets (cf.\ \citealt{oort51,whipple78}) and supports the idea that this was indeed its first pass close to the Sun. It is also unlikely that the ice on ISON became rapidly depleted during the apparition since this is believed to occur gradually over many orbits, not in just a few months and while the comet was still at normal cometary distances, near $r_\mathrm{H}= 1$ AU. Furthermore, at a minimum of one quarter of the total surface area of the nucleus, the active area necessary to support water production is hardly a ``small'' region of the surface and would not be expected to produce the well defined CN features we observed.

We propose that the CN features were caused by a localized enhancement of activity from a single, near equatorial source region in addition to the ongoing activity across the sunward hemisphere. This model would naturally explain the variations in PA and bulk brightness of the two CN features we observed in November. Our modeling suggests that the Sun remained within $\sim$30$^\circ$ of ISON's pole until just days before perihelion (when the true anomally finally began to change rapidly) 
so regions near its equator only received sunlight obliquely and therefore heated up slowly. Thus, enhanced activity from the source region may have been initiated in late October when local temperatures (either at the surface or below it as the thermal wave penetrated) climbed sufficiently to trigger significant outgassing for the first time. This initiation of activity may have been violent enough to expose substantial portions of the comet's pristine interior and likely lead to enhanced activity at this location (local topography may have also played a role in shaping the outgassing into a ``jet,'' e.g., \citealt{crifo02}). If this caused the creation of a hole or occurred on the side of a sunward facing slope, some regions likely would have experienced more direct sunlight than the average surface at that latitude, helping vigorous outgassing to continue and exposing ever larger portions of the interior.

The chronological appearance of the features first in narrowband gas images and later in broadband and unfiltered images acquired after the $\sim$12$\times$ increase in both gas and dust from November 11 to 13 suggests that the active region creating the features was the source of subsequent outbursts of activity.
The initial activity was only vigorous enough for the features to be seen in narrowband gas images. The accompanying dust likely did not achieve sufficient S/N to appear in broadband images until
the large outburst November 11--13, when the rate of outgassing from this region increased significantly compared to the ambient outgassing from the rest of the surface. At this time the features brightened enough relative to the ambient coma to become visible in broadband and even unfiltered images. This was likely due to more extreme excavation of the interior around the active region, such as break off of additional chunks, caving in of a crevasse wall, or extensive cracking. The later appearance of the features in broadband images was not likely due to a change in the dust-to-gas ratio at deeper depths, since the dust and gas production rates went up in unison during the outburst.

Note that the massive increase in gas and dust production that began November 11 was unlikely to have been due only to the release of one or a few large chunks that remained intact, as these would not have had nearly enough surface area to significantly affect the production rates. Instead, the increase likely required the release of a large number of small particles, either directly or through the rapid disintegration of larger particles, to efficiently increase the effective surface area within the coma. For example, the disintegration of a $\sim$10-m on a side cube into $\mu$m-sized icy grains would have provided sufficient surface area to achieve the increase in water production reported by \citet{cbet3711c}. In order to sustain this higher level of production, however, such mass loss would have needed to be ongoing, and the increasing production rates from November 11 to 23 would have required increasingly more material to be lost. Thus, the November 11 outburst may have triggered runaway mass loss that eventually led to the catastrophic failure of the nucleus.

\citet{cbet3715} suggested that the appearance of ``coma wings'' in their broadband images on November 14 signaled that the nucleus had recently split. Similar morphology has previously been noted in at least three split comets: C/1996 B2 Hyakutake \citep{harris97,rodionov98}, C/1999 S4 LINEAR \citep{farnham01a}, and C/2001 A2 LINEAR \citep{jehin02}; see also the review by \citet{boehnhardt04}. Of these, C/1999 S4 LINEAR is the best match for ISON, where two ``wings'' perpendicular to the comet-Sun line appeared $\sim$2 weeks prior to that comet's breakup. As with our interpetation of ISON's CN features, \citet{farnham01a} concluded that these wings were produced by a single active area near the equator of a rotating nucleus as opposed to the nucleus splitting at that time. Since ISON both lasted for several weeks after the first appearance of the features and continued to show similar features through at least November 18 (Boehnhardt et al.\footnote{http://isoncampaign.org/observation-logs\#20131118}), we contend that the first appearance of its features on or before November 1 was not caused by a final, catastrophic disruption of the nucleus at that time. 
Furthermore, the variation of the PAs and relative brightness of the features from night to night as well as the asymmetric nature of the features are all inconsistent with the ``wings'' seen in C/2001 A2 LINEAR that apparently originated between two fragments \citep{jehin02}.

If significant chunks of the nucleus were shed November 1 or later and remained intact, they would not likely have traveled far enough from the nucleus to be distinguishable in our images by November 12. Assuming a breakup on November 1.0 at a separation velocity of $\sim$1 m s$^{-1}$ \citep{sekanina82a,boehnhardt04} and ignoring gravitational or rocket effects, the maximum separation of two fragments was $\lesssim$1.5 arcsec by November 12, with projection effects likely decreasing the apparent separation in the plane of the sky. This is below typical seeing on these nights ($\sim$2 arcsec) and would, therefore, not have been detectable in our images.

\section{CONCLUSIONS}
\label{sec:conclusions}
By combining the observations presented herein with results from the community, a coherrent narrative for ISON's behavior is beginning to emerge. It appears that ISON was considerably smaller than original estimates based on the brightness at discovery suggested. This was likely due to the presence of a large cross section of dust and icy grains, many of which were probably expelled during an extended, CO-driven outburst as suggested by \citet{meech13}. These grains moved away from the nucleus very slowly, increasing ISON's apparent brightness and measured production rates for many months. The slow movement of such grains away from the nucleus and/or their gradual destruction helps explain a number of our observations during 2013: the flattening of the dust radial profile, {\afrho} remaining relatively constant, the shoulder in gas production rates, and the declining minimum active area. As a result, ISON underperformed most brightness projections until at least late October.

ISON's fortunes likely changed dramatically in late-October. Around this time we observed a pair of CN features that varied in PA and relative brightness from night to night. Our modeling suggests that these originated from a single source region located near the equator, and may have been triggered by the delayed heating of this region due to the Sun being nearly over the pole throughout ISON's approach. Beginning on November 11, ISON's dust and gas production rates increased rapidly \citep{cbet3711c} and similar morphological features first appeared in dust and broadband images. This suggests that an increasing fraction of activity was occuring from the source region, perhaps due to the opening of a sizeable fissure or the loss of one or more large chunks. Regardless of the mechanism, newly injected material must have quickly broken up into many small particles in order to supply the surface area needed to explain observed water production rates, which greatly exceeded the total surface area of an $R_N<0.6$ km nucleus \citep{cbet3720a}. Water production rates continued to increase until at least November 23.6 \citep{combi14}, implying that progressively more mass loss occured, possibly including fragmentation of the nucleus.

Possible explanations for the runaway mass loss include the loss of icy glue holding sections of the nucleus together, build up of subsurface pressure which was eventually released through one or more catastrophic outbursts, or splitting via rotational spin-up \citep{samarasinha13}. Significant mass loss was not likely triggered by tidal forces, which should not have become significant until the day of perihelion \citep{knight13b}. 
The recognition of this extreme mass loss helps explain why ISON's brightness and morphological behavior in {\it STEREO} and {\it SOHO} images in the days before perihelion \citep{knight14} was akin to that of the small ($R_N<50$ m) Kreutz sungrazing comets regularly observed by those telescopes being destroyed as they approach perihelion \citep{knight10d}. While ISON's brightness and gas production rates at larger heliocentric distances implied it was large enough to survive insolation despite its small perihelion distance \citep{knight13b}, the largest remaining fragment as it entered the {\it SOHO} fields of view was likely quite a bit smaller. Thus, ISON's destruction near perihelion should be seen not as a single catastrophic event, but as the culmination of a series of events which weakened and/or broke it up over several weeks leading up to perihelion.

\section*{ACKNOWLEDGMENTS}
We thank the anonymous referee for a helpful review. We gratefully acknowledge our various telescope operators at DCT: Stephen Levine, Alex Venetiou, Michael Sweaton, Jason Sanborn, Ron Winner, Lisa Foley, Susan Strosahl, and Heidi Larson for helping obtain successful observations during the early commissioning phase of the DCT. We thank Brian Skiff for obtaining the 42-in images on October 8--9, Larry Wasserman for scripting the 31-in robotic images, and Michaela Fendrock, Kevin Walsh, and Allison Bair for assistance in observing at DCT. We also thank BBC Horizons for showcasing the DCT during these observations, and Hermann Boehnhardt, Tony Farnham, and Michael A'Hearn for useful discussions.

These results made use of Lowell Observatory's Discovery Channel Telescope, supported by Lowell, Discovery Communications, Boston University, the University of Maryland, and the University of Toledo. M.M.K. is grateful for office space provided by the University of Maryland Department of Astronomy and Johns Hopkins University Applied Physics Laboratory while working on this project. The LMI instrument was funded by the National Science Foundation via grant AST-1005313. This research has been supported by NASA's Planetary Astronomy Program (Grants NNX09AB51G, NNX11AD95G, and NNX14AG81G).


\label{lastpage}

\end{singlespace}

\renewcommand{\baselinestretch}{0.78}
\renewcommand{\arraystretch}{1.0}

\begin{deluxetable}{lccrccccccc}  
\tabletypesize{\scriptsize}
\tablecolumns{11}
\tablewidth{0pt} 
\setlength{\tabcolsep}{0.05in}
\tablecaption{Imaging observations and geometric parameters for comet ISON (C/2012 S1).\,\tablenotemark{a}}
\tablehead{   
  \colhead{UT}&
  \colhead{UT}&
  \colhead{$\Delta$T}&
  \colhead{Tel\tablenotemark{b}}&
  \colhead{$r_\mathrm{H}$}&
  \colhead{$\Delta$}&
  \colhead{Phase}&
  \colhead{PA Sun}&
  \colhead{Filters}&
  \colhead{Conditions}\\
  \colhead{Date}&
  \colhead{Range}&
  \colhead{(days)}&
  \colhead{}&
  \colhead{(AU)}&
  \colhead{(AU)}&
  \colhead{($^\circ$)}&
  \colhead{($^\circ$)}&
  \colhead{}&
  \colhead{}
}
\startdata
Jan 12&02:57--03:11&$-$320.6&DCT&5.143&4.170&\phantom{0}1.8&\phantom{00}7.0&R&Clouds\\
Jan 13&02:39--02:44&$-$319.7&DCT&5.132&4.160&\phantom{0}1.9&359.7&R&Intermittent clouds\\
Mar 11&04:20--04:27&$-$262.6&DCT&4.500&4.080&12.1&277.3&R&Cirrus\\
Apr 4&02:48--02:56&$-$238.6&DCT&4.221&4.210&13.6&272.9&B,V,R&Cirrus\\
Apr 6&02:42--02:50&$-$236.7&DCT&4.198&4.221&13.7&272.7&B,V,R&Clouds\\
Apr 7&03:12--07:06&$-$235.6&42in&4.185&4.227&13.7&272.5&B,V,R,I,CN,BC&Photometric\\
Apr 8&03:03--07:02&$-$234.6&42in&4.173&4.232&13.7&272.4&B,V,R,I&Cirrus\\
May 1&03:14--03:18&$-$211.6&DCT&3.896&4.324&12.8&270.3&B,V,R&Cirrus\\
May 3&03:11--03:15&$-$209.6&DCT&3.871&4.329&12.6&270.1&B,V,R&Photometric\tablenotemark{c}\\
Jun 11&03:27--04:10&$-$170.6&42in&3.372&4.263&\phantom{0}7.4&265.4&V,R,I&Photometric\\
Sep 12&10:51--12:20&\phantom{0}$-$77.3&DCT&1.984&2.674&18.3&110.8&V,R,CN,BC&Photometric\\
Sep 13&11:30--12:22&\phantom{0}$-$76.3&DCT&1.967&2.646&18.7&110.8&V,R,CN&Clouds\\
Sep 30&10:40--12:23&\phantom{0}$-$59.3&DCT&1.661&2.166&26.3&111.5&V,R,OH,CN,C3,CO$+$,BC,C2,GC,RC&Photometric\\
Oct 1&11:05--12:34&\phantom{0}$-$58.3&DCT&1.642&2.136&26.9&111.5&V,R,CN,C3,BC,C2,GC&Photometric\\
Oct 2&11:05--12:30&\phantom{0}$-$57.3&DCT&1.623&2.107&27.4&111.6&B,V,R,I,CN,C3,BC,C2,GC&Photometric\\
Oct 3&10:59--12:30&\phantom{0}$-$56.3&DCT&1.604&2.077&27.9&111.6&B,V,R,I,CN,C3,CO$+$,BC,C2,GC&Photometric\\
Oct 4&11:09--12:26&\phantom{0}$-$55.3&DCT&1.584&2.047&28.5&111.7&B,V,R,I,CN,C3,BC,C2,GC,RC&Photometric\\
Oct 4&11:18--12:39&\phantom{0}$-$55.3&31in&1.584&2.047&28.5&111.7&R,CN&Photometric\\
Oct 5&11:04--11:59&\phantom{0}$-$54.3&42in&1.565&2.018&29.1&111.8&B,V,R,I,CN,C3,BC&Photometric\\
Oct 5&11:13--12:35&\phantom{0}$-$54.3&31in&1.565&2.018&29.1&111.8&R,CN&Photometric\\
Oct 6&11:00--11:53&\phantom{0}$-$53.3&42in&1.546&1.989&29.6&111.9&B,V,R,I,CN,C3,BC&Photometric\\
Oct 8&11:45--12:07&\phantom{0}$-$51.3&42in&1.506&1.928&30.8&112.0&R,CN&Cirrus\\
Oct 9&11:34--11:58&\phantom{0}$-$50.3&42in&1.487&1.898&31.4&112.1&R,CN&Cirrus\\
Oct 9&11:18--12:39&\phantom{0}$-$50.3&31in&1.487&1.898&31.4&112.1&R,CN&Cirrus\\
Oct 15&11:35--12:35&\phantom{0}$-$44.3&31in&1.365&1.717&35.5&112.6&R,CN&Photometric\\
Nov 1&11:23--12:18&\phantom{0}$-$27.3&31in&0.985&1.216&52.3&113.8&R,CN&Photometric\\
Nov 2&11:22--12:18&\phantom{0}$-$26.3&31in&0.960&1.189&53.7&113.9&R,CN&Photometric\\
Nov 4&11:04--12:26&\phantom{0}$-$24.3&31in&0.910&1.135&56.7&113.8&R,CN&Clouds\\
Nov 6&11:22--12:46&\phantom{0}$-$22.3&31in&0.859&1.084&60.0&113.8&R,CN&Photometric\\
Nov 7&11:22--12:44&\phantom{0}$-$21.3&31in&0.833&1.059&61.8&113.7&R,CN&Photometric\\
Nov 8&11:22--13:02&\phantom{0}$-$20.3&31in&0.806&1.035&63.7&113.5&R,CN&Cirrus\\
Nov 9&11:28--13:13&\phantom{0}$-$19.3&31in&0.778&1.012&65.6&113.3&R,CN&Photometric\\
Nov 10&11:24--13:05&\phantom{0}$-$18.3&31in&0.751&0.991&67.7&113.2&R,CN&Photometric\\
Nov 11&11:42--13:05&\phantom{0}$-$17.3&31in&0.723&0.970&69.9&112.9&R,CN&Photometric\\
Nov 12&11:44--13:08&\phantom{0}$-$16.2&31in&0.694&0.950&72.1&112.5&R,CN&Clouds\\
\enddata
\tablenotetext{a} {All parameters are given for the midpoint of each night's observations, and all images were obtained at Lowell Observatory in 2013.}
\tablenotetext{b} {DCT = Discovery Channel Telescope (4.3-m), 42in = Hall 42-in Telescope (1.1-m), 31in = 31-in Telescope (0.8-m).}
\tablenotetext{c} {A bright star was too close to the comet for {\afrho} to be measured reliably.}
\label{t:imaging_circ}
\end{deluxetable}

\begin{deluxetable}{lllcccccccc}  
\tabletypesize{\scriptsize}
\tablecolumns{11}
\tablewidth{0pt} 
\setlength{\tabcolsep}{0.03in}
\tablecaption{Photometry observing circumstances and fluorescence efficiencies for comet ISON (C/2012 S1).\,\tablenotemark{a}}
\tablehead{   
  \multicolumn{2}{c}{UT Date}&
  \colhead{$\Delta$T}&
  \colhead{$r_\mathrm{H}$}&
  \colhead{$\Delta$}&
  \colhead{Phase}&
  \colhead{Phase}&
  \colhead{$\dot{r}_\mathrm{H}$}&
  \multicolumn{3}{c}{log $L/N$ (erg s$^{-1}$ molecule$^{-1}$)}\\
  \cmidrule(){9-11}
  \colhead{}&
  \colhead{}&
  \colhead{(day)}&
  \colhead{(AU)}&
  \colhead{(AU)}&
  \colhead{($^\circ$)}&
  \colhead{Adj.\,\tablenotemark{b}}&
  \colhead{(km s$^{-1}$)}&
  \colhead{OH}&
  \colhead{NH}&
  \colhead{CN}
}
\startdata
Mar&\phantom{0}5.2&$-$268.6&4.554&4.039&\phantom{0}11.3&0.18&$-$19.6&...&...&$-$12.379\\
Apr&\phantom{0}7.2&$-$235.6&4.185&4.227&\phantom{0}13.7&0.21&$-$20.6&...&...&...\\
May&\phantom{0}4.2&$-$208.6&3.847&4.321&\phantom{0}12.5&0.20&$-$21.3&...&...&$-$12.369\\
Sep& 12.5&\phantom{0}$-$77.3&1.981&2.670&\phantom{0}18.3&0.27&$-$29.7&...&...&$-$12.508\\
Sep& 14.5&\phantom{0}$-$75.3&1.947&2.616&\phantom{0}19.1&0.28&$-$30.0&$-$14.507&$-$13.108&$-$12.499\\
Sep& 30.5&\phantom{0}$-$59.3&1.659&2.164&\phantom{0}26.4&0.35&$-$32.5&...&...&$-$12.446\\
Oct&\phantom{0}1.5&\phantom{0}$-$58.3&1.640&2.134&\phantom{0}26.9&0.36&$-$32.7&...&...&$-$12.448\\
Oct&\phantom{0}2.5&\phantom{0}$-$57.3&1.621&2.105&\phantom{0}27.4&0.36&$-$32.9&...&...&$-$12.450\\
Oct&\phantom{0}3.5&\phantom{0}$-$56.3&1.602&2.075&\phantom{0}28.0&0.37&$-$33.1&...&...&$-$12.453\\
Oct&\phantom{0}4.5&\phantom{0}$-$55.3&1.583&2.046&\phantom{0}28.5&0.37&$-$33.3&$-$14.454&$-$13.118&$-$12.456\\
Oct&\phantom{0}5.5&\phantom{0}$-$54.3&1.564&2.017&\phantom{0}29.1&0.38&$-$33.5&...&...&$-$12.460\\
Oct&\phantom{0}6.5&\phantom{0}$-$53.3&1.544&1.987&\phantom{0}29.6&0.38&$-$33.7&...&...&$-$12.463\\
Oct& 15.5&\phantom{0}$-$44.3&1.363&1.716&\phantom{0}35.5&0.42&$-$35.8&...&...&$-$12.474\\
Nov&\phantom{0}1.5&\phantom{0}$-$27.3&0.984&1.216&\phantom{0}52.4&0.47&$-$42.1&...&...&$-$12.396\\
Nov&\phantom{0}2.5&\phantom{0}$-$26.3&0.960&1.188&\phantom{0}53.8&0.47&$-$42.6&...&...&$-$12.405\\
Nov&\phantom{0}6.5&\phantom{0}$-$22.3&0.858&1.084&\phantom{0}60.1&0.46&$-$45.1&...&...&$-$12.437\\
Nov&\phantom{0}7.5&\phantom{0}$-$21.3&0.832&1.059&\phantom{0}61.8&0.46&$-$45.8&...&...&$-$12.439\\
Nov&\phantom{0}9.5&\phantom{0}$-$19.3&0.778&1.012&\phantom{0}65.7&0.44&$-$47.3&...&...&$-$12.432\\
Nov& 10.5&\phantom{0}$-$18.3&0.751&0.990&\phantom{0}67.7&0.44&$-$48.1&...&...&$-$12.427\\
Nov& 11.5&\phantom{0}$-$17.3&0.723&0.970&\phantom{0}69.8&0.42&$-$49.1&...&...&$-$12.423\\

\enddata
\tablenotetext{a} {All parameters are given for the midpoint of each night's observations, and all images were obtained at Lowell Observatory in 2013.}
\tablenotetext{b} {Adjustment to 0\deg\ phase angle to log({\afrho}) values based on assumed phase function (see text).}
\label{t:phot_circ}
\end{deluxetable}

\begin{deluxetable}{lllcccccccccccccccc}  
\tabletypesize{\scriptsize}
\tablecolumns{19}
\tablewidth{0pt} 
\setlength{\tabcolsep}{0.02in}
\tablecaption{Photometric fluxes and aperture abundances for comet ISON (C/2012 S1).\tablenotemark{a}}
\tablehead{   
  \multicolumn{2}{c}{UT Date}&
  \colhead{Tel\tablenotemark{b}}&
  \multicolumn{2}{c}{Aperture}&
  \multicolumn{5}{c}{log Emission Band Flux}&
  \multicolumn{4}{c}{log Continuum Flux\tablenotemark{c,d}}&
  \multicolumn{5}{c}{log $M$($\rho$)}\\
  \cmidrule(lr){4-5}
  \colhead{}&
  \colhead{}&
  \colhead{}&
  \colhead{Size}&
  \colhead{log $\rho$}&
  \multicolumn{5}{c}{(erg cm$^{-2}$ s$^{-1}$)}&
  \multicolumn{4}{c}{(erg cm$^{-2}$ s$^{-1}$ \AA$^{-1}$)}&
  \multicolumn{5}{c}{(molecule)}\\
  \cmidrule(lr){6-10}
  \cmidrule(lr){11-14}
  \cmidrule(lr){15-19}
  \colhead{}&
  \colhead{}&
  \colhead{}&
  \colhead{(arcsec)}&
  \colhead{(km)}&
  \colhead{OH}&
  \colhead{NH}&
  \colhead{CN}&
  \colhead{C$_3$}&
  \colhead{C$_2$}&
  \colhead{UV}&
  \colhead{Blue}&
  \colhead{Green}&
  \colhead{Red}&
  \colhead{OH}&
  \colhead{NH}&
  \colhead{CN}&
  \colhead{C$_3$}&
  \colhead{C$_2$}

}
\startdata
Mar&\phantom{0}5.2&42in\tablenotemark{e}&\phantom{0}97.2&5.15&...&...&$-$13.35&...&...&...&$-$14.98&...&...&...&...&29.01&...&...\\
Apr&\phantom{0}7.2&42in\tablenotemark{e}&\phantom{0}16.4&4.40&...&...&...&...&...&...&...&...&-15.08&...&...&...&...&...\\
May&\phantom{0}4.2&42in\tablenotemark{e}&\phantom{0}62.4&4.99&...&...&$-$13.20&...&...&...&$-$14.69&...&...&...&...&29.06&...&...\\
Sep& 12.5&DCT&\phantom{0}25.9&4.40&...&...&$-$12.23&...&...&...&$-$14.12&...&-14.30&...&...&29.17&...&...\\
Sep& 14.5&42in\tablenotemark{e}&\phantom{0}97.2&4.96&$-$10.53&$-$11.95&$-$11.39&$-$11.29&$-$11.17&{\it und}&$-$13.69&$-$13.78&...&32.84&30.02&29.97&29.58&30.04\\
Sep& 30.5&DCT&\phantom{0}32.0&4.40&...&...&$-$11.57&$-$11.95&$-$11.70&...&$-$13.86&...&-14.04&...&...&29.44&28.60&29.21\\
Oct&\phantom{0}1.5&DCT&\phantom{0}32.5&4.40&...&...&$-$11.71&$-$11.75&$-$11.61&...&$-$13.94&...&-14.03&...&...&29.27&28.78&29.27\\
Oct&\phantom{0}2.5&DCT&\phantom{0}32.9&4.40&...&...&$-$11.76&$-$11.74&$-$11.00&...&$-$13.93&...&-14.02&...&...&29.21&28.78&29.86\\
Oct&\phantom{0}3.5&DCT&\phantom{0}33.4&4.40&...&...&$-$11.68&$-$11.61&$-$11.51&...&$-$13.92&...&-13.98&...&...&29.27&28.88&29.33\\
Oct&\phantom{0}4.5&42in\tablenotemark{e}&155.9&5.06&\phantom{0}$-$9.79&$-$11.05&$-$10.68&$-$10.97&$-$10.45&$-$13.44&$-$13.44&$-$13.44&...&33.14&30.53&30.24&29.50&30.37\\
Oct&\phantom{0}4.5&42in\tablenotemark{e}&\phantom{0}97.2&4.86&$-$10.04&$-$11.39&$-$10.95&$-$11.19&$-$10.75&$-$13.71&$-$13.50&$-$13.48&...&32.88&30.19&29.98&29.28&30.07\\
Oct&\phantom{0}4.5&42in\tablenotemark{e}&\phantom{0}48.6&4.56&$-$10.53&$-$11.87&$-$11.37&$-$11.42&$-$11.15&$-$14.02&$-$13.75&$-$13.77&...&32.39&29.72&29.56&29.05&29.66\\
Oct&\phantom{0}4.5&DCT&\phantom{0}33.9&4.40&...&...&$-$11.67&$-$11.67&$-$11.54&...&$-$13.88&$-$13.90&-13.98&...&...&29.25&28.80&29.28\\
Oct&\phantom{0}4.5&31in&\phantom{0}33.9&4.40&...&...&$-$11.49&...&...&...&...&...&-13.96&...&...&29.44&...&...\\
Oct&\phantom{0}5.5&42in&\phantom{0}34.4&4.40&...&...&$-$11.61&$-$11.63&...&...&$-$13.89&...&-13.96&...&...&29.29&28.82&...\\
Oct&\phantom{0}5.5&31in&\phantom{0}34.4&4.40&...&...&$-$11.50&...&...&...&...&...&-13.92&...&...&29.40&...&...\\
Oct&\phantom{0}6.5&42in&\phantom{0}34.9&4.40&...&...&$-$11.54&$-$11.56&...&...&$-$13.89&...&-13.95&...&...&29.34&28.86&...\\
Oct& 15.5&31in&\phantom{0}40.4&4.40&...&...&$-$11.10&...&...&...&...&...&-13.75&...&...&29.55&...&...\\
Nov&\phantom{0}1.5&31in&\phantom{0}57.1&4.40&...&...&$-$10.20&...&...&...&...&...&-13.22&...&...&29.80&...&...\\
Nov&\phantom{0}2.5&31in&\phantom{0}58.4&4.40&...&...&$-$10.15&...&...&...&...&...&-13.09&...&...&29.82&...&...\\
Nov&\phantom{0}6.5&31in&\phantom{0}64.0&4.40&...&...&\phantom{0}$-$9.86&...&...&...&...&...&-12.99&...&...&29.96&...&...\\
Nov&\phantom{0}7.5&31in&\phantom{0}65.5&4.40&...&...&\phantom{0}$-$9.74&...&...&...&...&...&-12.92&...&...&30.04&...&...\\
Nov&\phantom{0}9.5&31in&\phantom{0}68.6&4.40&...&...&\phantom{0}$-$9.58&...&...&...&...&...&-12.82&...&...&30.09&...&...\\
Nov& 10.5&31in&\phantom{0}70.0&4.40&...&...&\phantom{0}$-$9.49&...&...&...&...&...&-12.78&...&...&30.12&...&...\\
Nov& 11.5&31in&\phantom{0}71.5&4.40&...&...&\phantom{0}$-$9.36&...&...&...&...&...&-12.68&...&...&30.19&...&...\\

\enddata
\tablenotetext{a} {All parameters are given for the midpoint of each night's observations, and all images were obtained at Lowell Observatory in 2013.}
\tablenotetext{b} {DCT = Discovery Channel Telescope (4.3-m), 42in = Hall 42-in Telescope (1.1-m), 31in = 31-in Telescope (0.8-m).}
\tablenotetext{c} {Continuum filter wavelengths: UV = 3448 \AA; blue = 4450 \AA; green = 5260 \AA; red = $\sim$6500 \AA\ (Cousins R).}
\tablenotetext{d} {{\it ``und''} stands for ``undefined'' and means the continuum flux was measured but was less than 0.}
\tablenotetext{e} {Data obtained with photometer. All other data were acquired with a CCD.}
\label{t:phot_flux}
\end{deluxetable}

\begin{deluxetable}{lllccccccccccccc}  
\tabletypesize{\scriptsize}
\tablecolumns{16}
\tablewidth{0pt} 
\setlength{\tabcolsep}{0.02in}
\tablecaption{Photometric production rates for comet ISON (C/2012 S1).\tablenotemark{a}}
\tablehead{   
  \multicolumn{2}{c}{UT Date}&
  \colhead{$\Delta$T}&
  \colhead{Tel\tablenotemark{b}}&
  \colhead{log $r_\mathrm{H}$}&
  \colhead{log $\rho$}&
  \multicolumn{5}{c}{log $Q$\tablenotemark{c}\phantom{00}(molecules s$^{-1}$)}&
  \multicolumn{4}{c}{log $A$($\theta$)$f\rho$\tablenotemark{c,d,e}\phantom{000}(cm)}&
  \colhead{log $Q$}\\
  \cmidrule(lr){7-11}
  \cmidrule(lr){12-15}
  \cmidrule(){16-16}
  \colhead{}&
  \colhead{}&
  \colhead{(day)}&
  \colhead{}&
  \colhead{(AU)}&
  \colhead{(km)}&
  \colhead{OH}&
  \colhead{NH}&
  \colhead{CN}&
  \colhead{C$_3$}&
  \colhead{C$_2$}&
  \colhead{UV}&
  \colhead{Blue}&
  \colhead{Green}&
  \colhead{Red}&
  \colhead{H$_2$O}

}
\startdata
Mar&\phantom{0}5.2&$-$268.6&42in\tablenotemark{f}&\phantom{$-$}0.658&5.15&...{\tiny\phantom{.00}}&...{\tiny\phantom{.00}}&24.11{\tiny\phantom{.}.12}&...{\tiny\phantom{.00}}&...{\tiny\phantom{.00}}&...{\tiny\phantom{.00}}&2.06{\tiny\phantom{.}.13}&...{\tiny\phantom{.00}}&...{\tiny\phantom{.00}}&...\\
Apr&\phantom{0}7.2&$-$265.6&42in\tablenotemark{f}&\phantom{$-$}0.622&4.40&...{\tiny\phantom{.00}}&...{\tiny\phantom{.00}}&...{\tiny\phantom{.00}}&...{\tiny\phantom{.00}}&...{\tiny\phantom{.00}}&...{\tiny\phantom{.00}}&...{\tiny\phantom{.00}}&...{\tiny\phantom{.00}}&2.88{\tiny\phantom{.}.08}&...\\
May&\phantom{0}4.2&$-$208.6&42in\tablenotemark{f}&\phantom{$-$}0.585&4.99&...{\tiny\phantom{.00}}&...{\tiny\phantom{.00}}&24.33{\tiny\phantom{.}.16}&...{\tiny\phantom{.00}}&...{\tiny\phantom{.00}}&...{\tiny\phantom{.00}}&2.43{\tiny\phantom{.}.13}&...{\tiny\phantom{.00}}&...{\tiny\phantom{.00}}&...\\
Sep& 12.5&\phantom{0}$-$77.3&DCT&\phantom{$-$}0.297&4.40&...{\tiny\phantom{.00}}&...{\tiny\phantom{.00}}&25.05{\tiny\phantom{.}.08}&...{\tiny\phantom{.00}}&...{\tiny\phantom{.00}}&...{\tiny\phantom{.00}}&2.59{\tiny\phantom{.}.08}&...{\tiny\phantom{.00}}&2.62{\tiny\phantom{.}.08}&...\\
Sep& 14.5&\phantom{0}$-$75.3&42in\tablenotemark{f}&\phantom{$-$}0.289&4.96&28.04{\tiny\phantom{.}.03}&25.41{\tiny\phantom{.}.19}&25.04{\tiny\phantom{.}.06}&24.75{\tiny\phantom{.}.15}&25.30{\tiny\phantom{.}.04}&{\it und}&2.42{\tiny\phantom{.}.08}&2.36{\tiny\phantom{.}.09}&...{\tiny\phantom{.00}}&28.03\\
Sep& 30.5&\phantom{0}$-$59.3&DCT&\phantom{$-$}0.220&4.40&...{\tiny\phantom{.00}}&...{\tiny\phantom{.00}}&25.23{\tiny\phantom{.}.08}&24.25{\tiny\phantom{.}.08}&25.20{\tiny\phantom{.}.08}&...{\tiny\phantom{.00}}&2.51{\tiny\phantom{.}.08}&...{\tiny\phantom{.00}}&2.54{\tiny\phantom{.}.08}&...\\
Oct&\phantom{0}1.5&\phantom{0}$-$58.3&DCT&\phantom{$-$}0.215&4.40&...{\tiny\phantom{.00}}&...{\tiny\phantom{.00}}&25.06{\tiny\phantom{.}.08}&24.43{\tiny\phantom{.}.08}&25.25{\tiny\phantom{.}.08}&...{\tiny\phantom{.00}}&2.41{\tiny\phantom{.}.08}&...{\tiny\phantom{.00}}&2.54{\tiny\phantom{.}.08}&...\\
Oct&\phantom{0}2.5&\phantom{0}$-$57.3&DCT&\phantom{$-$}0.210&4.40&...{\tiny\phantom{.00}}&...{\tiny\phantom{.00}}&24.99{\tiny\phantom{.}.08}&24.42{\tiny\phantom{.}.08}&25.84{\tiny\phantom{.}.08}&...{\tiny\phantom{.00}}&2.40{\tiny\phantom{.}.08}&...{\tiny\phantom{.00}}&2.51{\tiny\phantom{.}.08}&...\\
Oct&\phantom{0}3.5&\phantom{0}$-$56.3&DCT&\phantom{$-$}0.205&4.40&...{\tiny\phantom{.00}}&...{\tiny\phantom{.00}}&25.04{\tiny\phantom{.}.08}&24.53{\tiny\phantom{.}.08}&25.30{\tiny\phantom{.}.08}&...{\tiny\phantom{.00}}&2.39{\tiny\phantom{.}.08}&...{\tiny\phantom{.00}}&2.53{\tiny\phantom{.}.08}&...\\
Oct&\phantom{0}4.5&\phantom{0}$-$55.3&42in\tablenotemark{f}&\phantom{$-$}0.199&5.06&28.17{\tiny\phantom{.}.01}&25.71{\tiny\phantom{.}.03}&25.17{\tiny\phantom{.}.01}&24.73{\tiny\phantom{.}.06}&25.50{\tiny\phantom{.}.01}&2.51{\tiny\phantom{.}.07}&2.18{\tiny\phantom{.}.06}&2.21{\tiny\phantom{.}.06}&...{\tiny\phantom{.00}}&28.20\\
Oct&\phantom{0}4.5&\phantom{0}$-$55.3&42in\tablenotemark{f}&\phantom{$-$}0.199&4.86&28.16{\tiny\phantom{.}.01}&25.66{\tiny\phantom{.}.03}&25.14{\tiny\phantom{.}.01}&24.59{\tiny\phantom{.}.06}&25.42{\tiny\phantom{.}.01}&2.45{\tiny\phantom{.}.07}&2.33{\tiny\phantom{.}.04}&2.37{\tiny\phantom{.}.04}&...{\tiny\phantom{.00}}&28.20\\
Oct&\phantom{0}4.5&\phantom{0}$-$55.3&42in\tablenotemark{f}&\phantom{$-$}0.199&4.56&28.09{\tiny\phantom{.}.01}&25.64{\tiny\phantom{.}.04}&25.11{\tiny\phantom{.}.01}&24.56{\tiny\phantom{.}.05}&25.40{\tiny\phantom{.}.02}&2.43{\tiny\phantom{.}.07}&2.37{\tiny\phantom{.}.04}&2.38{\tiny\phantom{.}.04}&...{\tiny\phantom{.00}}&28.13\\
Oct&\phantom{0}4.5&\phantom{0}$-$55.3&DCT&\phantom{$-$}0.199&4.40&...{\tiny\phantom{.00}}&...{\tiny\phantom{.00}}&25.02{\tiny\phantom{.}.08}&24.44{\tiny\phantom{.}.08}&25.25{\tiny\phantom{.}.08}&...{\tiny\phantom{.00}}&2.41{\tiny\phantom{.}.08}&2.41{\tiny\phantom{.}.08}&2.50{\tiny\phantom{.}.08}&...\\
Oct&\phantom{0}4.5&\phantom{0}$-$55.3&31in&\phantom{$-$}0.199&4.40&...{\tiny\phantom{.00}}&...{\tiny\phantom{.00}}&25.21{\tiny\phantom{.}.08}&...{\tiny\phantom{.00}}&...{\tiny\phantom{.00}}&...{\tiny\phantom{.00}}&...{\tiny\phantom{.00}}&...{\tiny\phantom{.00}}&2.53{\tiny\phantom{.}.08}&...\\
Oct&\phantom{0}5.5&\phantom{0}$-$54.3&42in&\phantom{$-$}0.194&4.40&...{\tiny\phantom{.00}}&...{\tiny\phantom{.00}}&25.06{\tiny\phantom{.}.08}&24.46{\tiny\phantom{.}.08}&...{\tiny\phantom{.00}}&...{\tiny\phantom{.00}}&2.38{\tiny\phantom{.}.08}&...{\tiny\phantom{.00}}&2.50{\tiny\phantom{.}.08}&...\\
Oct&\phantom{0}5.5&\phantom{0}$-$54.3&31in&\phantom{$-$}0.194&4.40&...{\tiny\phantom{.00}}&...{\tiny\phantom{.00}}&25.17{\tiny\phantom{.}.08}&...{\tiny\phantom{.00}}&...{\tiny\phantom{.00}}&...{\tiny\phantom{.00}}&...{\tiny\phantom{.00}}&...{\tiny\phantom{.00}}&2.55{\tiny\phantom{.}.08}&...\\
Oct&\phantom{0}6.5&\phantom{0}$-$53.3&42in&\phantom{$-$}0.189&4.40&...{\tiny\phantom{.00}}&...{\tiny\phantom{.00}}&25.10{\tiny\phantom{.}.08}&24.50{\tiny\phantom{.}.08}&...{\tiny\phantom{.00}}&...{\tiny\phantom{.00}}&2.35{\tiny\phantom{.}.08}&...{\tiny\phantom{.00}}&2.49{\tiny\phantom{.}.08}&...\\
Oct& 15.5&\phantom{0}$-$44.3&31in&\phantom{$-$}0.134&4.40&...{\tiny\phantom{.00}}&...{\tiny\phantom{.00}}&25.26{\tiny\phantom{.}.08}&...{\tiny\phantom{.00}}&...{\tiny\phantom{.00}}&...{\tiny\phantom{.00}}&...{\tiny\phantom{.00}}&...{\tiny\phantom{.00}}&2.45{\tiny\phantom{.}.08}&...\\
Nov&\phantom{0}1.5&\phantom{0}$-$27.3&31in&$-$0.007&4.40&...{\tiny\phantom{.00}}&...{\tiny\phantom{.00}}&25.42{\tiny\phantom{.}.08}&...{\tiny\phantom{.00}}&...{\tiny\phantom{.00}}&...{\tiny\phantom{.00}}&...{\tiny\phantom{.00}}&...{\tiny\phantom{.00}}&2.40{\tiny\phantom{.}.08}&...\\
Nov&\phantom{0}2.5&\phantom{0}$-$26.3&31in&$-$0.018&4.40&...{\tiny\phantom{.00}}&...{\tiny\phantom{.00}}&25.44{\tiny\phantom{.}.08}&...{\tiny\phantom{.00}}&...{\tiny\phantom{.00}}&...{\tiny\phantom{.00}}&...{\tiny\phantom{.00}}&...{\tiny\phantom{.00}}&2.39{\tiny\phantom{.}.08}&...\\
Nov&\phantom{0}6.5&\phantom{0}$-$22.3&31in&$-$0.067&4.40&...{\tiny\phantom{.00}}&...{\tiny\phantom{.00}}&25.56{\tiny\phantom{.}.08}&...{\tiny\phantom{.00}}&...{\tiny\phantom{.00}}&...{\tiny\phantom{.00}}&...{\tiny\phantom{.00}}&...{\tiny\phantom{.00}}&2.41{\tiny\phantom{.}.08}&...\\
Nov&\phantom{0}7.5&\phantom{0}$-$21.3&31in&$-$0.080&4.40&...{\tiny\phantom{.00}}&...{\tiny\phantom{.00}}&25.64{\tiny\phantom{.}.08}&...{\tiny\phantom{.00}}&...{\tiny\phantom{.00}}&...{\tiny\phantom{.00}}&...{\tiny\phantom{.00}}&...{\tiny\phantom{.00}}&2.44{\tiny\phantom{.}.08}&...\\
Nov&\phantom{0}9.5&\phantom{0}$-$19.3&31in&$-$0.109&4.40&...{\tiny\phantom{.00}}&...{\tiny\phantom{.00}}&25.69{\tiny\phantom{.}.08}&...{\tiny\phantom{.00}}&...{\tiny\phantom{.00}}&...{\tiny\phantom{.00}}&...{\tiny\phantom{.00}}&...{\tiny\phantom{.00}}&2.43{\tiny\phantom{.}.08}&...\\
Nov& 10.5&\phantom{0}$-$18.3&31in&$-$0.124&4.40&...{\tiny\phantom{.00}}&...{\tiny\phantom{.00}}&25.72{\tiny\phantom{.}.08}&...{\tiny\phantom{.00}}&...{\tiny\phantom{.00}}&...{\tiny\phantom{.00}}&...{\tiny\phantom{.00}}&...{\tiny\phantom{.00}}&2.43{\tiny\phantom{.}.08}&...\\
Nov& 11.5&\phantom{0}$-$17.3&31in&$-$0.141&4.40&...{\tiny\phantom{.00}}&...{\tiny\phantom{.00}}&25.79{\tiny\phantom{.}.08}&...{\tiny\phantom{.00}}&...{\tiny\phantom{.00}}&...{\tiny\phantom{.00}}&...{\tiny\phantom{.00}}&...{\tiny\phantom{.00}}&2.48{\tiny\phantom{.}.08}&...\\

\enddata
\tablenotetext{a} {All parameters are given for the midpoint of each night's observations, and all images were obtained at Lowell Observatory in 2013.}
\tablenotetext{b} {DCT = Discovery Channel Telescope (4.3-m), 42in = Hall 42-in Telescope (1.1-m), 31in = 31-in Telescope (0.8-m).}
\tablenotetext{c} {Production rates followed by the upper, i.e. the positive, uncertainty. The ``+'' and ``$-$'' uncertainties are equal as percentages, but unequal in log-space; the ``$-$'' values can be computed.}
\tablenotetext{d} {Continuum filter wavelengths: UV = 3448 \AA; blue = 4450 \AA; green = 5260 \AA; red = $\sim$6500 \AA\ (Cousins R).}
\tablenotetext{e} {{\it ``und''} stands for ``undefined'' and means the continuum flux was measured but was less than 0.}\label{t:phot_rates}
\tablenotetext{f} {Data obtained with photometer. All other data were acquired with a CCD.}
\label{t:phot_rates}
\end{deluxetable}

\begin{deluxetable}{lcc}  
\tabletypesize{\scriptsize}
\tablecolumns{3}
\tablewidth{0pt} 
\setlength{\tabcolsep}{0.05in}
\tablecaption{Position angles of CN features.} 
\tablehead{   
  \colhead{Date}&
  \colhead{North}&
  \colhead{South}\\
  \colhead{}&
  \colhead{PA ($^\circ$)\tablenotemark{a}}&
  \colhead{PA ($^\circ$)\tablenotemark{a}}
}
\startdata
2013 Nov 1&\phantom{$3$}30&190\\
2013 Nov 2&\phantom{$3$}50&200\\
2013 Nov 4&\phantom{$3$}40&210\\
2013 Nov 6&\phantom{$3$}30&210\\
2013 Nov 7&\phantom{$3$}20&220\\
2013 Nov 8&\phantom{$3$}40&210\\
2013 Nov 9&\phantom{$3$}40&210\\
2013 Nov 10&\phantom{$3$}50&210\\
2013 Nov 11&\phantom{$3$}30&220\\
2013 Nov 12&340&210\\
\enddata
\tablenotetext{a} {The uncertainty of all PAs is estimated to be $\pm$10$^\circ$.}
\label{t:cn_pas}
\label{lasttable}
\end{deluxetable}

\clearpage

\renewcommand{\baselinestretch}{0.8}

\begin{figure}
  \centering
  \includegraphics[width=88mm]{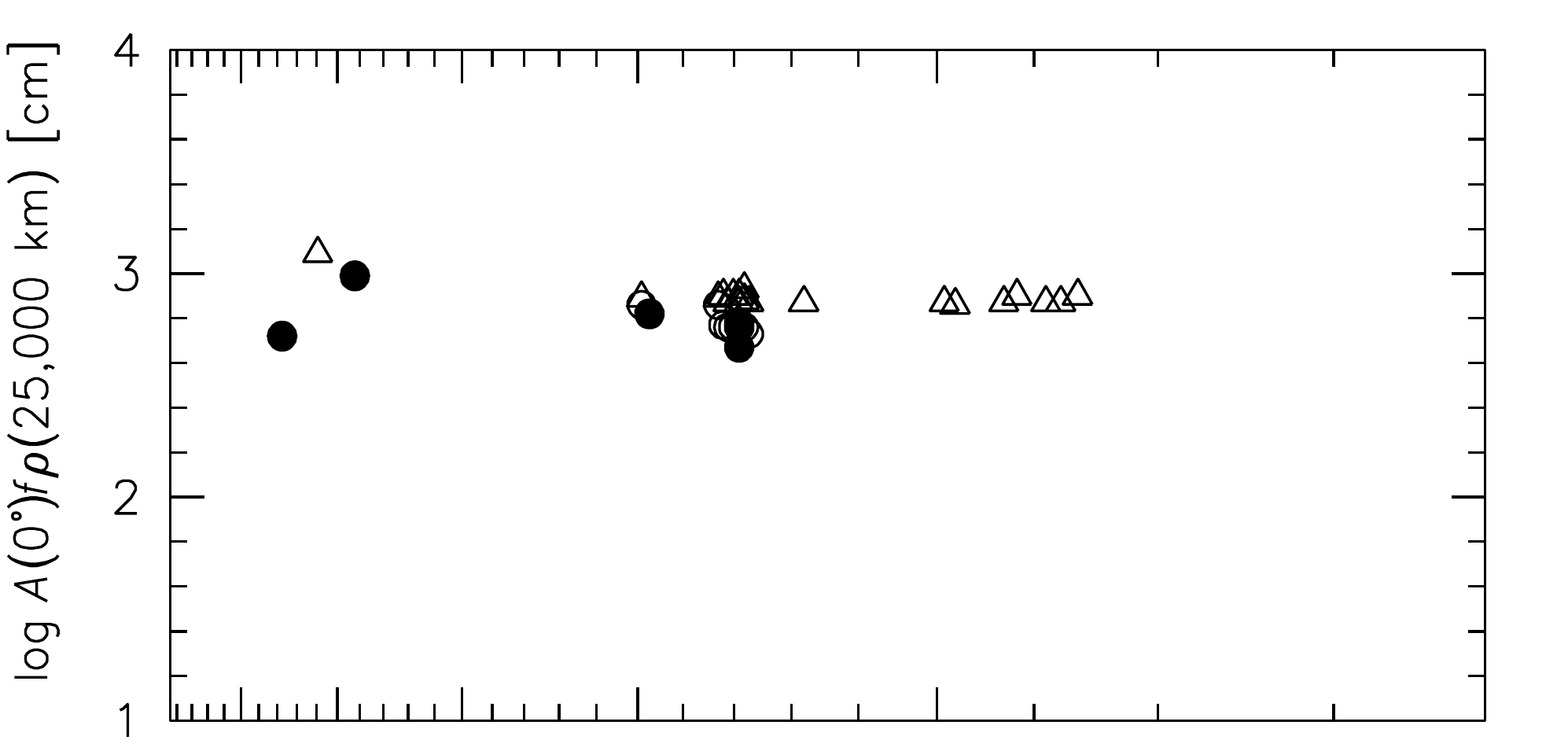}
  \includegraphics[width=88mm]{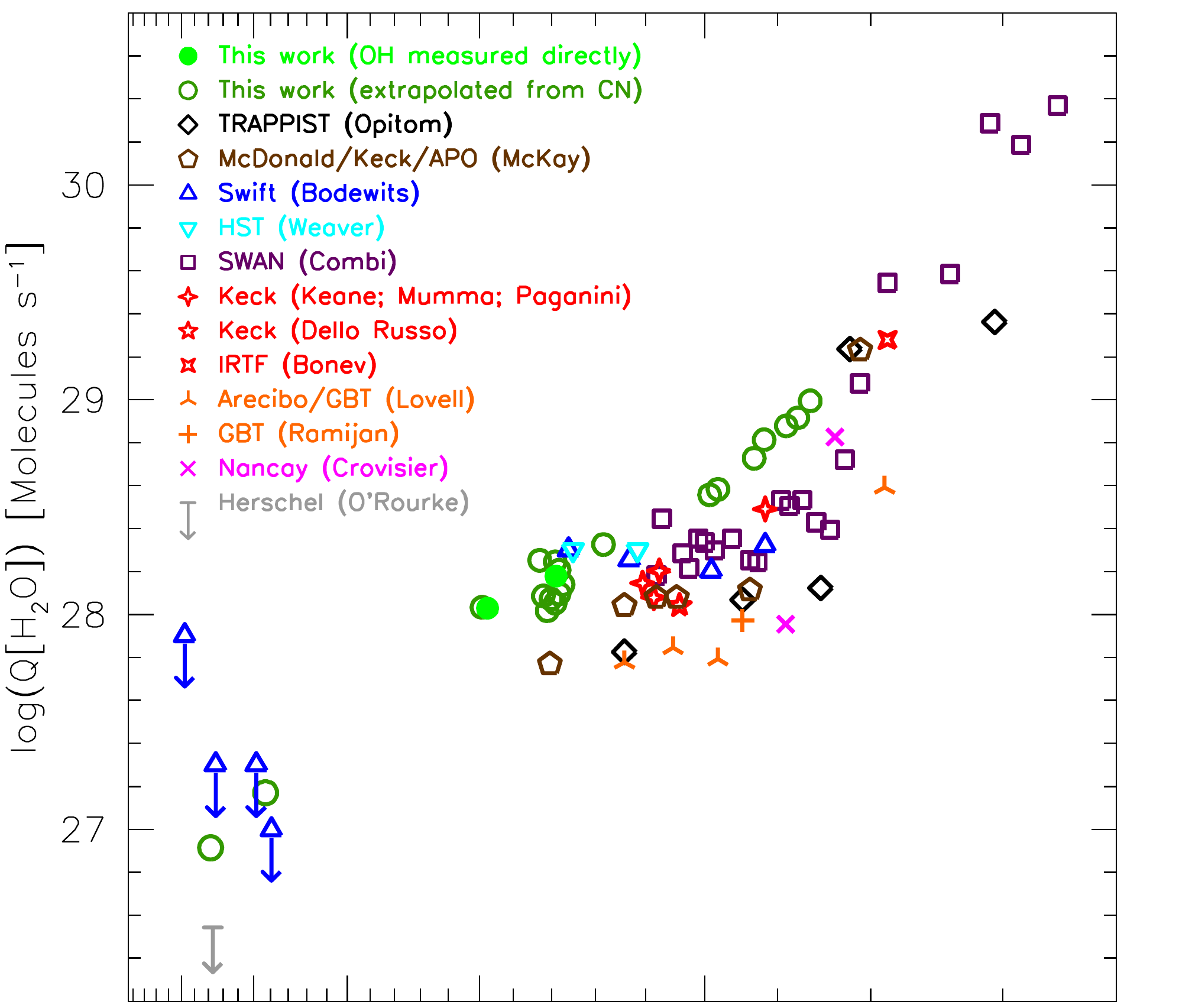}
  \includegraphics[width=88mm]{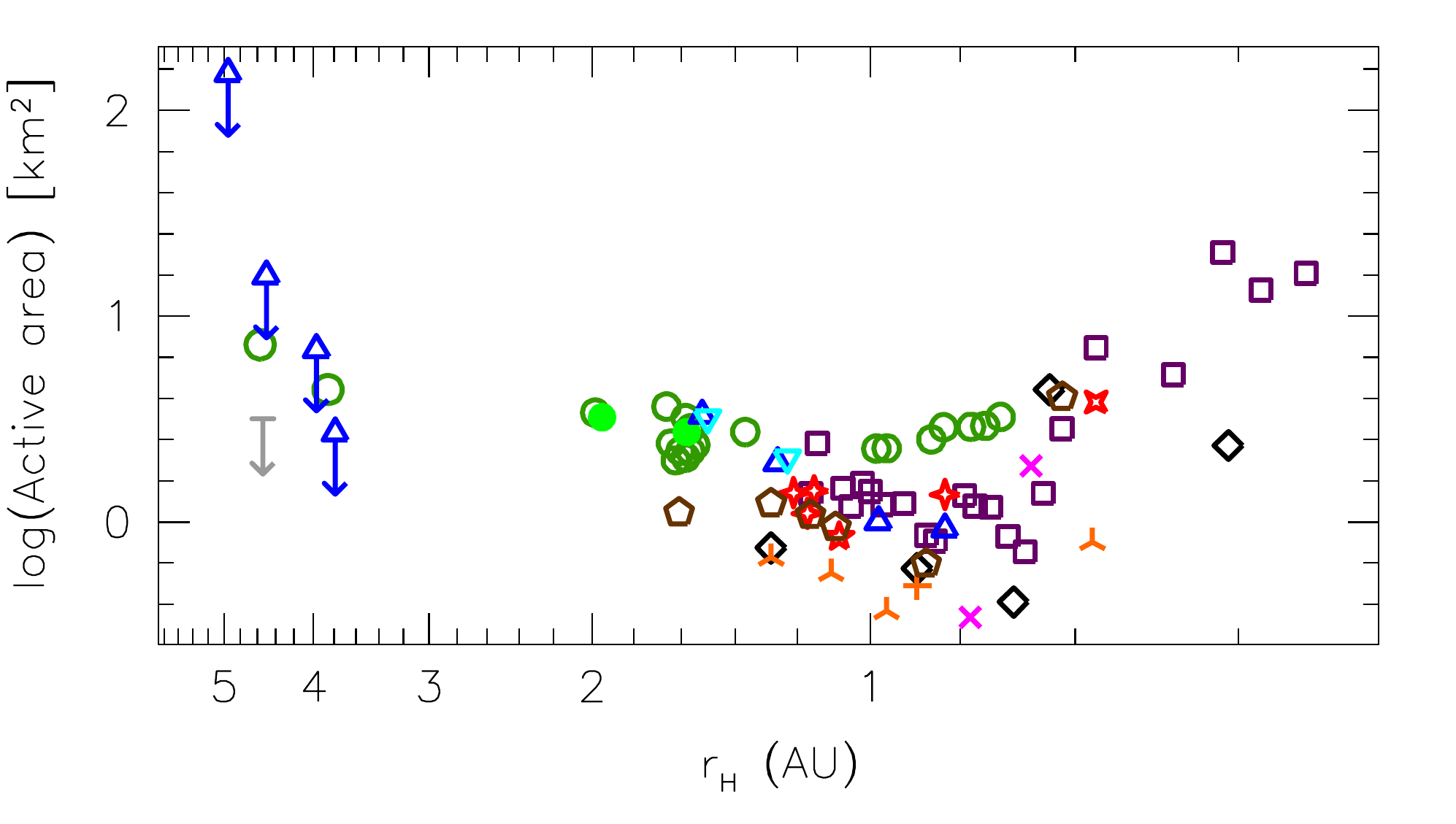}
  \label{fig:water_rates}
\end{figure}

\begin{figure}
  \centering
  \caption[Log of dust production rates, normalized for both aperture size and phase angle (top panel), log of water production rate (middle panel), and active area (bottom panel) versus heliocentric distance]{Log of dust production rates, normalized for both aperture size and phase angle versus heliocentric distance (plotted with a logarithmic scale in all panels; top panel), log of water production rate versus heliocentric distance (middle panel), and log of active area versus heliocentric distance (bottom panel). All data in the top panel are from this paper and the symbols are as defined in Figure~\ref{fig:photometry}. All measurements derived from CCD images used an aperture of 25,000 km while all photometer measurements were normalized to $\rho$ = 25,000 km by extrapolation from the curves shown in Figure~\ref{fig:afrho}. 
The data in the bottom two panels are from this paper and compiled from the literature. The symbols in the bottom two panels are the same and are defined in the key of the middle panel. Error bars are omitted for clarity while upper limits are indicated by a downward arrow extending from the point. Our values are based on OH and/or extrapolated from CN (see text). 
Other OH-based results include {\it HST}/STIS \citep{cbet3680}, {\it Swift} \citep{cbet3718a}, TRAPPIST \citep{cbet3693a,cbet3711c}, GBT \citep{cbet3693c}, Arecibo/GBT (Lovell, private communication 2014), and Nan\c{c}ay \citep{cbet3711a}. \citet{mckay14} used OH on two nights and [OI] for all others. Grand-daughter hydrogen was measured with {\it SOHO}/SWAN by \citet{combi14}. Direct water measurements were made with Keck/NIRSPEC by \citet{iauc9261a}, \citet{iauc9261b}, \citet{iauc9263b} (plotted together with one symbol) and \citet{cbet3686}, and with IRTF/CSHELL \citep{cbet3720b}; upper limits were acquired with {\it Herschel}/HIFI (\citealt{orourke13}; they present two model-dependent upper limits and we plot the lower).
In cases where the authors reported a range of measurements and/or dates, 
we plot the average of the measurement range at the midpoint of the date range. We converted reported $Q$(OH) to $Q$(H$_2$O) by 1.361$r_\mathrm{H}^{-0.5}$$Q$(OH) \citep{cochran93,schleicher98}. Minimum active areas were derived from H$_2$O production rates following the methodology of \citet{cowan79} and assuming the subsolar case.
}
  \label{fig:water_rates}
\end{figure}

\end{document}